\documentclass[twocolumn,showpacs,aps,prx,longbibliography,superscriptaddress]{revtex4-1}
\usepackage{graphicx,amsmath,amssymb}
\usepackage{dcolumn,fancyhdr}
\usepackage{bm,natbib,mathrsfs}
\usepackage{times}
\usepackage{txfonts,palatino}
\usepackage{epstopdf}
\usepackage{latexsym}
\usepackage{epsfig}
\usepackage{bbold}

\newcommand{\ket}[1]{\left\vert#1\right\rangle}

\newcommand{\bra}[1]{\left\langle#1\right\vert}

	\newcommand{\tr}[1]{\textrm{Tr} \left[ {#1} \right]} 

	\newcommand{\e}[1]{e^{ {#1}}} 

\begin{document}

\title{Assessing the non-equilibrium thermodynamics in a quenched quantum many-body system via single projective measurements}

\author{L. Fusco} 
\affiliation{Centre for Theoretical Atomic, Molecular and Optical Physics, School of Mathematics and Physics, Queen's University, Belfast BT7 1NN, United Kingdom}
\author{S. Pigeon}
\affiliation{Centre for Theoretical Atomic, Molecular and Optical Physics, School of Mathematics and Physics, Queen's University, Belfast BT7 1NN, United Kingdom}
\author{T. Apollaro}
\affiliation{Dipartimento di Fisica, Universita' della Calabria, 87036 Arcavacata di Rende (CS), Italy}
\affiliation{Centre for Theoretical Atomic, Molecular and Optical Physics, School of Mathematics and Physics, Queen's University, Belfast BT7 1NN, United Kingdom}
\author{A. Xuereb}
\affiliation{Centre for Theoretical Atomic, Molecular and Optical Physics, School of Mathematics and Physics, Queen's University, Belfast BT7 1NN, United Kingdom}
\affiliation{Department of Physics, University of Malta, Msida MSD 2080, Malta}
\author{L. Mazzola}
\affiliation{Centre for Theoretical Atomic, Molecular and Optical Physics, School of Mathematics and Physics, Queen's University, Belfast BT7 1NN, United Kingdom}
\author{M. Campisi}
\affiliation{NEST, Scuola Normale Superiore \& Istituto di Nanoscienze-CNR, I-56126 Pisa, Italy}
\author{ A. Ferraro}
\affiliation{Centre for Theoretical Atomic, Molecular and Optical Physics, School of Mathematics and Physics, Queen's University, Belfast BT7 1NN, United Kingdom}
\author{M. Paternostro}
\affiliation{Centre for Theoretical Atomic, Molecular and Optical Physics, School of Mathematics and Physics, Queen's University, Belfast BT7 1NN, United Kingdom}
\author{ G. De Chiara}
\affiliation{Centre for Theoretical Atomic, Molecular and Optical Physics, School of Mathematics and Physics, Queen's University, Belfast BT7 1NN, United Kingdom}
\date{\today}

\begin{abstract}
We analyse the nature of the statistics of the work done on or by a quantum many-body system brought out of equilibrium. We show that, for the sudden quench and for an initial state which commutes with the initial Hamiltonian, it is possible to retrieve the whole non-equilibrium thermodynamics via single projective measurements of observables. We highlight in a physically clear way the qualitative implications for the statistics of work coming from considering processes described by operators that either commute or do not commute with the unperturbed Hamiltonian of a given system. We consider a quantum many-body system and derive an expression that allows us to give a physical interpretation, for a thermal initial state, to all of the cumulants of the work in the case of quenched operators commuting with the unperturbed Hamiltonian. In the commuting case the observables that we need to measure have an intuitive physical meaning. Conversely, in the non-commuting case we show that, although it is possible to operate fully within the single-measurement framework irrespectively of the size of the quench, some difficulties are faced in providing a clear-cut physical interpretation to the cumulants. This circumstance makes the study of the physics of the system non-trivial and highlights the non-intuitive phenomenology of the emergence of thermodynamics from the fully quantum microscopic description. We illustrate our ideas with the example of the Ising model in a transverse field showing the interesting behavior of the high-order statistical moments of the work distribution for a generic thermal state and linking them to the critical nature of the model itself.

\end{abstract}

\maketitle

A considerable amount of effort has been made, recently, on the study of the statistics of work in quantum systems subject to a process~\cite{lutz,varie,Abah,Ngo,CPmaps,Oxf,KavanJohn,JohnMauroKavan,Sindona,SindonaPRL,Def1,Burada,Def2,QJ,Morillo}. One of the interests in this area lies in the possibility to predict the exact value taken by thermodynamically relevant quantities (such as work, free-energy variations, and entropy) by analysing the features of explicitly finite-time, out-of-equilibrium dynamics. Such possibility, which is embodied by elegant fluctuation theorems~\cite{esposito,campisi,seifert,Tasaki,Crooks,Jarzynski}, has been demonstrated experimentally in both the classical and quantum mechanical scenarios~\cite{Liphardt, Collin, Douarche, Toyabe, Saira, Batalhao}. The increasing level of control, at the quantum level, of simple systems consisting of a few quantum particles~\cite{Batalhao,Huber,Heyl,Pekola,Dorner, CampisiNew} thus makes this an exciting time to investigate questions related to the thermodynamics of explicitly non-equilibrium processes. 

A quite natural step forward in this direction is given by the extensions of such studies to the quantum many-body domain, whose rich physics and phenomenology would offer unmatched possibilities to explore thermodynamically relevant questions from a genuine quantum mechanical viewpoint. Interesting first attempts in this sense, for both spin and harmonic systems, have been reported recently in Refs.~\cite{Silva,Dorner,Joshi,Smacchia,dudu,carlisle,tony}. Yet, notwithstanding the key contributions that such endeavors embody in the quest for the establishment of a bridge between thermodynamics and the physics of quantum systems, there is a pressing need for a deeper comprehension of the links between quantum criticality and the statistical mechanics of work and entropy arising from out-of-equilibrium processes. 

In fact, much remains to be understood of the way thermodynamics {\it emerges} from quantum critical phenomena in extended quantum systems and, in turn, how we can signal the occurrence of criticality by looking at thermodynamic quantities. This is exactly the goal of this paper, which aims at providing a physical interpretation of the statistical moments of the work distribution following a non-adiabatic transformation on a quantum many-body system based on the commutation relations among the various parts of the system's Hamiltonian. At this aim we show that there are conditions under which it is possible to retreive the thermodynamic quantities by doing single projective measurements.\\

We analyse in detail the full statistics of the work distribution in a quantum many-body system. We report explicit expressions for all the moments and cumulants of the work distribution in the case of a sudden quench of a Hamiltonian parameter. In particular we analyse the case of a system subject to a sudden switch of an external magnetic field, and show that, as long as the quenching process is described by an operator that commutes with the unperturbed part of the Hamiltonian, the cumulants of the work distribution have a fairly intuitive physical interpretation. 

In order to address the case of an experimentally accessible observable, we focus on the magnetization of a many-body system and link analytically its cumulants to higher-order susceptibilities. This allows us to make explicit statements on the possibility of observing signatures of quantum criticality in the cumulants of the work distribution. 
We provide an interesting paradigm of our investigation by studying the work distribution for the Ising model in a transverse field. 
Our study paves the way to the revelation of quantum critical effects via the assessment of the full statistics of work, and strengthens the interesting connections between the emergence of non-equilibrium thermodynamics and macroscopic properties in many-body physics, whose investigation is currently only in its infancy~\cite{Silva,Dorner,Joshi,Smacchia,dudu, carlisle}.

The remainder of this paper is organised as follows: In Sec.~\ref{qfr}, we briefly review the formalism of quantum fluctuation relations, introducing the explicit form of the probability distribution of  work and its characteristic function for any generic initial state of the system. In Sec.~\ref{sudden} we discuss the validity of the sudden quench assumption. Sec.~\ref{interpretation}, provides the physical interpretation of high-order cumulants of such distribution based on the commutativity (or lack thereof) of the 
Hamiltonian of a quantum many-body system before and after a quantum process. In Sec.~\ref{model}, we use the transverse Ising model to illustrate the key findings of our theoretical analysis, demonstrating that the work statistics indeed brings about information on the critical nature of the model at hand by showcasing a neat singularity at low temperature that is progressively smeared out as soon as thermal fluctuations start dominating. In Sec.~\ref{conc}, we summarise our findings and discuss possible open routes.

\section{Quantum fluctuation relations: a brief review}
\label{qfr}

 Here we give a brief summary of the formalism that will be used throughout this work. We consider a process undergone by a system $S$ and described by a Hamiltonian $\hat{\cal H}(\lambda_t)$ depending on a {\it work parameter} $\lambda_t$, which is assumed to be externally controlled.
At $t=0^-$, $S$ is initialised in a generic quantum state $\hat{\rho}_0$. At $t=0^+$,  while keeping the system isolated, we perform a {\it process} consisting of the change of $\lambda_t$ to its final value $\lambda_\tau$.
It is convenient to decompose the Hamiltonians connected by the process as
\begin{equation}
\hat{\cal H}(\lambda_0)=\sum_{n,i} E_n(\lambda_0) \ket{n^{(i)}(\lambda_0)}\bra{n^{(i)}(\lambda_0)}
\end{equation}
and
\begin{equation}
\hat{\cal H}(\lambda_\tau)=\sum_{m,j} E_m(\lambda_\tau) \ket{m^{(j)}(\lambda_{\tau})} \bra{m^{(j)}(\lambda_{\tau})},
\end{equation}
where $\left\{E_n(\lambda_0),\ket{n^{(i)}(\lambda_0)}\right\}$ $\left[\left\{E_m(\lambda_{\tau}),\ket{m^{(j)}(\lambda_{\tau})}\right\}\right]$ is the $n^\textrm{th}$ [$m^\textrm{th}$] eigenvalue-eigenstate pair of the initial [final] Hamiltonian, and $i$ and $j$ are quantum numbers labelling the possible degeneracy of the eigenvalues. The corresponding work distribution can be written as~\cite{Tasaki,Morillo2}
\begin{equation}
P(W):=\sum_{n,m} p^0_n\;  p^\tau_{m \vert n} \delta\left[W-(E_m(\lambda_{\tau})-E_n(\lambda_0))\right].
\label{defwork}
\end{equation}
Here, we have introduced the probability with which the particular eigenvalue $E_n(\lambda_0)$ is observed in the first measurement performed over the system
\begin{equation}
p^0_n=\tr{\hat{P}_n(\lambda_0) \hat{\rho}_0},
\end{equation}
where
\begin{equation}
\hat{P}_n(\lambda_0)=\sum_{i} \ket{n^{(i)}(\lambda_0)}\bra{n^{(i)}(\lambda_0)}
\end{equation}
is the projector onto the eigensubspace of the eigenvalue $E_n(\lambda_0)$. The conditional probability of observing the eigenvalue $E_m(\lambda_{\tau})$ at time $t=\tau$, after the observation of $E_n(\lambda_0)$ at time $t=0$, is given by
\begin{equation}
p_{m|n}^{\tau}=\tr{\hat{P}_m(\lambda_{\tau})\hat{U}_{\tau}\hat{\rho}_n\hat{U}_{\tau}^{\dagger}},
\end{equation}
where
\begin{equation}
\hat{\rho}_n=\frac{\hat{P}_n(\lambda_0)\hat{\rho}_0 \hat{P}_n(\lambda_0)}{p_n^0}
\end{equation}
is the state in which the system is found immediately after the first measurement, and $\hat{U}_{\tau}$ is the evolution operator describing the process.

$P(W)$ encompasses the statistics of the initial state (given by $p_n^0$) and the fluctuations arising from quantum measurement statistics (given by $p^\tau_{m \vert n}$). It is possible to show that the characteristic function of work, for a generic initial state $\hat{\rho}_0$, can be written as \cite{Morillo2}
\begin{equation}
\chi(u,\tau)=\int\!{dW}\e{iuW}P(W)=\tr{U^\dag_\tau\e{iu\hat{\cal H}(\lambda_\tau)}\hat U_\tau\e{-iu\hat{\cal H}(\lambda_0)}\hat{\rho}_{0}'},
\label{characteristicfunction}
\end{equation}
where
\begin{equation}
\hat{\rho}_0'=\sum_n \hat{P}_n(\lambda_0) \hat{\rho}_0\hat{P}_n(\lambda_0)
\end{equation}
is the state of the system projected onto the eigenbasis of the Hamiltonian $\hat{{\cal H}}(\lambda_0)$, hereafter called the \emph{initial projected state}. In particular, the following relation holds
\begin{equation}
\label{equivalence}
\hat{\rho}_0'=\hat{\rho}_0 \iff \left[ \hat{\rho}_0 , \hat{{\cal H}}(\lambda_0) \right]=0.
\end{equation}
If we restrict our attention to the case in which the initial state is a thermal state $\hat{\rho}_{G}(\lambda_0)={\e{-\beta\hat{\cal H}(\lambda_0)}}/{{\cal Z}(\lambda_0)}$ \footnote{In what follows we indicate physical quantities related to thermal states with a subscript $G$ (Gibbs). In Sec.~\ref{model} instead we will be interested only in the thermal state so that we will drop the subscript.} (where $\mathcal{Z}(\lambda)=\textrm{Tr}{\e{-\beta\hat{\cal H}(\lambda)}}$ is the partition function) at inverse temperature $\beta$, then the relation in Eq.~\eqref{equivalence} trivially holds, and from Eq.~\eqref{characteristicfunction} the Jarzynski equality~\cite{Jarzynski} is found as 
\begin{equation}
\chi_G(i\beta,\tau)= \langle \e{-\beta W} \rangle_G=\e{-\beta \Delta F}.
\end{equation}
The characteristic function is also crucial for the establishment of the Tasaki-Crooks relation $\Delta F=-(1/\beta)\ln[\chi_G(u,\tau)/\chi_G'(i\beta-u,\tau)]$~\cite{Crooks,Tasaki} with $\chi_G'(v,\tau)$ the characteristic function of the backward process obtained taking $\lambda_\tau\to\lambda_0$ and evolving $\hat{\rho}_G(\lambda_{\tau})$ backward. Here $\Delta{F}$ is the net change in the equilibrium free-energy of $S$. This demonstrates the central role played by the characteristic function in determining the equilibrium properties of a system. 

\section{On the validity of the sudden-quench assumption}
\label{sudden}
Most of the analysis made in this paper makes use of sudden quench processes. This process is a sudden switch of the work parameter from the initial value $\lambda_0$ to the final $\lambda_{\tau}$, performed after detaching the system from the thermal reservoir that determines its initial equilibrium state, if the initial state is actually a thermal state. 
Regardless of the specific nature of the process that we consider, a sudden quench encompasses a very interesting case to study due to its highly non-adiabatic nature. Our aim here is to provide a semi-quantitative criterion that any quench in a general quantum many-body system should satisfy in order to be rightly considered as sudden.

In such a transformation, the Hamiltonian changes so quickly that the state of the system freezes. 
The time taken to change the Hamiltonian should thus be much shorter than the typical time-scale of the evolution of the system. Despite the quench being a non-perturbative process in general, it is possible to determine the probability for a state of the system to change, while changing the Hamiltonian, in a perturbative treatment with respect to the time-scale required for such change to occur~\cite{Migdal}.

We can consider the general Hamiltonian 
\begin{equation}
\hat{{\mathcal H}}(\lambda_t) =
\left\{
\begin{array}{c c}
\hat{{\mathcal H}}(\lambda_0) & t\le0, \\
\hat{{\mathcal H}}(\lambda_{\tau}) &  t>\tau.\\
\end{array}
\right.
\end{equation}
For $0<t<\tau$ (with $\tau$ small enough) the Hamiltonian is explicitly changing from $\hat{{\mathcal H}}(\lambda_0)$ to $\hat{{\mathcal H}}(\lambda_{\tau})$. For simplicity we will consider the case of a pure initial state and estimate the probability that the system makes a transition to a different state while we change the Hamiltonian.

We rewrite the Hamiltonian as
\begin{equation}
\hat{{\mathcal H}}(\lambda_t)=\hat{\mathcal H}(\lambda_{\tau})+\hat{V}(t),
\end{equation}
where $\hat{V}(t)=\hat{{\mathcal H}}(\lambda_t)-\hat{{\mathcal H}}(\lambda_{\tau})$. With this simple rewriting of the Hamiltonian, for $\tau$ small enough, we can treat the term $\hat{V}(t)$, at every time $t >0$, perturbatively with respect to $\hat{{\mathcal H}}(\lambda_{\tau})$. Let us decompose the actual state of the system in terms of the eigenstates of $\hat{\mathcal H}(\lambda_{\tau})$ as
\begin{equation}
\ket{\psi (t)}=\sum_m a_m(t) \ket{m(\lambda_{\tau})},~~~a_m(t)\in\mathbb{C}.
\end{equation}
The time evolution of the coefficients $a_m(t)$ of such decomposition is given by
\begin{equation}
i \frac{d}{dt}a_n(t)=\sum_m V_{nm}(t)e^{-i\left( E_m'-E_n' \right)t}a_m(t),
\end{equation}
where $V_{nm}(t)=\bra{n(\lambda_{\tau})}\hat{V}(t)\ket{m(\lambda_{\tau})}$. Formally integrating such expression, we get
\begin{equation}
a_m(t)=a_n(0)-i\sum_m \int_0^t V_{nm}(t') a_m(t')e^{-i\left( E_m'-E_n' \right)t'}dt'.
\label{coeffpert}
\end{equation}
The perturbative parameter is considered to be $V_{nm} \tau \ll 1$, so that the potential term $V_{nm}$ does not need to be small, but its period of action does correspondingly. If we assume that $\left(E'_n-E'_m\right)\tau \ll 1$, then the exponential term in Eq.~\eqref{coeffpert} is approximately unity.
 The zeroth order approximation coefficient, in the parameter $V_{nm} \tau$, is given by $a_m^{(0)}(t)=a_m(0)$ and we can substitute it into the right-hand side of Eq.~\eqref{coeffpert} to get the first order perturbation term
\begin{equation}
a_n^{(1)}(t)=-i \sum_m \int_0^t \bra{n(\lambda_{\tau})} \hat{V}(t')\ket{m(\lambda_\tau)}a_m(0)dt'.
\end{equation}
If the initial state of the system is an eigenstate of $\hat{\mathcal H}(\lambda_0)$, let us call it $\ket{i}=\sum_m a_m(0) \ket{m(\lambda_{\tau})}$, then the probability of the state making a transition to the state $\ket{n_{\tau}}$ is given by
\begin{equation}
\begin{aligned}
\left|a_n^{(1)}(t)\right|^2&=\left|\int_0^t \bra{n(\lambda_{\tau})} \hat{V}(t')\ket{i}dt'\right|^2\\
&=\left|\int_0^t \bra{n(\lambda_{\tau})} \left( \hat{\mathcal H}(\lambda_{t'})-\hat{\mathcal H}(\lambda_{\tau})  \right)\ket{i} dt'\right|^2.\\
\end{aligned}
\end{equation}
This probability is clearly model-dependent, so that it is not possible to make a general statement about the sudden nature of the quench if we do not specify the context explicitly. If, for example, we consider a Hamiltonian model of the form
\begin{equation}
\hat{\mathcal H}(\lambda_t)=\hat{A}+\lambda(t) \hat{B},
\end{equation}
the requirement of the sudden quench is translated into
\begin{equation}
\left| a_n^{(1)}(t)\right|^2=\left|\bra{n(\lambda_{\tau})}\hat{B} \ket{i}\right|^2\left|\int_0^t \left( \lambda_{\tau}-\lambda(t') \right)dt' \right|^2\ll1\hspace{0.5cm}\forall n,
\end{equation}
which is both dependent on the quenched operator matrix elements and on the way we change the parameter $\lambda$. If, for example, we assume a linear quench $\lambda(t)=\lambda_{\tau} t/\tau$ in the short time interval $0<t<\tau$, the upper bound on $\tau$ is given by
\begin{equation}
\tau\ll \frac{2}{\lambda_{\tau} \left|\bra{n(\lambda_{\tau})}\hat{B} \ket{i}\right|}.
\label{suddencondition}
\end{equation}
We thus need to find $\ket{n(\lambda_{\tau})}$ such that $\left|\bra{n(\lambda_{\tau})}\hat{B} \ket{i}\right|$ is maximum and then change our Hamiltonian in a time $\tau$ given by Eq.~\eqref{suddencondition}. This is a strong condition though, as if it is fulfilled the state is not changing at all. If for example the quenched operator is not limited in the norm instead, then there is no sharp condition of applicability of the sudden quench. In this case a more qualitative consideration can be made in relation to the typical time scale of evolution of the state. Indeed, we could for example take the characteristic time of evolution of a relevant observable for the system and give a looser condition requiring that $\tau$ would be smaller than this time scale.

It is worth stressing that the condition for a sudden change in a quantum many-body system has been met experimentally in different contents ranging from ultra-cold atomic systems to trapped ions~\cite{experiments}.

\section{Interpretation of the statistical moments of the work distribution for a sudden quench} 
\label{interpretation}

The work characteristic function for a sudden quench reads 
\begin{equation}
\chi(u,\tau)=\tr{\e{iu\hat{\cal H}(\lambda_\tau)}\e{-iu\hat{\cal H}(\lambda_0)}\hat{\rho}_0'}.
\label{characteristicfunctionsudden}
\end{equation}
Eq.~\eqref{characteristicfunctionsudden} can be expanded in power series as 
\begin{equation}
\chi(u,\tau)=\sum_{n=0}^{\infty}\frac{(i u)^n}{n!}~\langle W^n \rangle,
\label{seriesmoments}
\end{equation}
where we have introduced the statistical moments
\begin{equation}
\langle W^n \rangle =(-i)^n \left.\partial^n_u\,\chi(u,\tau)\right|_{u=0}.
\label{moments}
\end{equation}
By considering $\log\bigl[\chi(u,\tau)\bigr]$, on the other hand, one can introduce the cumulants $K_n$ as
\begin{equation}
K_n =(-i)^n \left.\partial^n_u\,\log\bigl[\chi(u,\tau)\bigr]\right|_{u=0}.
\label{cumulants}
\end{equation}
For a generic $\hat{\rho}_0'$ the first and second moments of the work are
\begin{equation}
\label{firstmomentGeneric}
\langle W \rangle =\tr{\left(\hat{\cal H}(\lambda_\tau)-\hat{\cal H}(\lambda_0)\right)\hat{\rho}_0'},
\end{equation}
\begin{equation}
\label{secondmomentGeneric}
\langle W^2 \rangle =\tr{\left(\hat{\cal H}(\lambda_\tau)^2-2\hat{\cal H}(\lambda_\tau)\hat{\cal H}(\lambda_0)+\hat{\cal H}(\lambda_0)^2\right)\hat{\rho}_0'}.
\end{equation}
However, the initial projected state $\hat{\rho}_0'$ commutes with the initial Hamiltonian because it is the diagonal part of $\hat{\rho}_0$ in the eigenbasis of $\hat{{\cal H}}(\lambda_0)$, so that by using the cyclic permutation invariance of the trace we are able to link both the first and second moments of the characteristic function to the net variation of the Hamiltonian of the system as
\begin{equation}
\label{firstsecondmoment}
\langle W^j \rangle =\tr{\left(\hat{\cal H}(\lambda_\tau)-\hat{\cal H}(\lambda_0)\right)^j\hat{\rho}_0'},
\end{equation}
for $j=1,2$ and for any $\hat{\rho}_0'$.\\
Eqs.~(\ref{firstmomentGeneric})-(\ref{firstsecondmoment}) hold also for the case of a general time dependent protocol as long as the final Hamiltonian  $\hat{\cal H}(\lambda_{\tau})$ is replaced by the Heisenberg representation of the Hamiltonian $\hat{\cal H}_H(\lambda_{\tau})=\hat{U}_\tau^{\dagger} \hat{H}(\lambda_{\tau}) \hat{U}_\tau$ \cite{nolte}.

In order to gain a physical insight on these two quantities we consider the case of a many-body system, e.g. a system composed of $N$ spin-$1/2$ particles whose Hamiltonian we cast into the form
\begin{equation}
\hat{{\cal H}}(\lambda_t)=\hat{{\cal H}}_{ss}-\lambda_t \sum_{i=1}^N \hat{\sigma}_i^z
\label{egmodel}
\end{equation}
with $\hat{{\cal H}}_{ss}$ a generic \emph{spin-spin} interaction term and $\hat\sigma^k_i~(k=x,y,z)$ the $k$-Pauli spin operator. The second term in Eq.~(\ref{egmodel}) is proportional to the $z$-magnetisation $\hat{M}_z=\sum_{i=1}^N \hat{\sigma}_i^z$ of the collection of spins. In what follows, we will make clear distinction between the case $[\hat{\cal H}_{ss},\hat M_z]=0$ and the case $[\hat{\cal H}_{ss},\hat M_z]\neq0$, hereafter respectively called \emph{commuting case} and \emph{non-commuting case}. In Eq.~\eqref{egmodel} $\lambda_t$ is the strength of an external magnetic field, given in units of the characteristic spin-spin interaction rate that characterises $\hat{\cal H}_{ss}$, and embodies the work parameter of our quenched process. The process we are interested in is the sudden change of the magnetic field by the amount $\Delta \lambda=\lambda_{\tau}-\lambda_0.$\\
The first moments of the work distribution are obtained by plugging Eq. \eqref{egmodel} in Eqs. \eqref{firstmomentGeneric}, \eqref{secondmomentGeneric} and \eqref{firstsecondmoment}. For the first moment, for any $\hat{\rho}_0'$, we find
\begin{equation}
\langle W\rangle=-\Delta \lambda \langle \hat{M}_z\rangle,
\label{firstwork}
\end{equation}
where the average is taken over $\hat{\rho}_0'$. This is in agreement with our intuition, as the process at hand consists of changing the Hamiltonian of the system through a varying magnetic field. Then, the change of the energy in the system, i.e. the work done on it, is expected to be proportional to the magnetization. For the second moment we find, again for any $\hat{\rho}_0'$,
\begin{equation}
\label{secondworkGeneric}
\langle W^2\rangle=(\Delta \lambda)^2 \langle \hat{M}_z^2\rangle.
\end{equation}

\subsection{Commuting case: $[\hat{\cal H}_{ss},\hat M_z]=0$}
We now focus our attention on the physical meaning of the cumulants of the work distribution. One of the reasons for looking deeper at this quantities is given by the fact that higher-order cumulants (such as the fourth-order and higher ones) have been at the center of substantial studies on the characterization of the quantum phase transition occurring in many body systems in light of their sensitivity to the details of the corresponding distribution~\cite{binder,binder2}.
In order to give a physical interpretation of the moments or cumulants of the work distribution, let us consider separately the commuting and non commuting cases.

Let us start with the commuting case $[\hat{\cal H}_{ss},\hat M_z]=0$. Eq.~\eqref{secondworkGeneric} tells us also that the variance of the work distribution is proportional to the variance of the magnetization distribution. In the commuting case the variance of the longitudinal magnetization, $[\Delta \hat M_z^2]_G= \langle \hat{M}_z^2\rangle_G- \langle \hat{M}_z\rangle_G^2 $, evaluated over a thermal state, is proportional to the magnetic susceptibility \cite{schwabl}
\begin{equation}
\chi_{M} := \frac{\partial \langle\hat{M}_z \rangle_G}{\partial \lambda_0}=\beta [\Delta \hat{M}_z^2]_G.
\label{susceptibility}
\end{equation}
Thus the thermal state is a useful special case. Indeed, for such a state and any given commuting model (i.e. any $\hat{{\cal H}}_{ss}$ such that $[\hat{\cal H}_{ss},\hat M_z]=0$), it is straightforward to gather a physical intuition of the meaning of the first two cumulants of the work distribution. These are given by the magnetization and the magnetic susceptibility of the initial thermal state, respectively. These embody two of the most relevant and well-studied quantities in the physics of a magnetic system~\cite{MattisNolting}.
Moreover we found out that Eq.~\eqref{susceptibility} is actually a specific case of a more general relation between the derivatives of the average magnetization and the higher cumulants of its distribution. Specifically, in Appendix A we show that, by introducing the proper moment-generating function for the observable $\hat{M}_z$ when the system is prepared in a thermal state, i.e.
\begin{equation}
[G (v,\lambda_0)]_G= \frac{\tr{e^{i v \hat{M}_z} e^{-\beta(\hat{\mathcal H}_{ss}-\lambda_0 \hat{M}_z)}}}{\mathcal{Z}(\lambda_0)},
\label{caraG}
\end{equation}
and the associated cumulants
\begin{equation}
[C_{n}(\lambda_0)]_G=\frac{1}{i^n}\frac{\partial^n \log [G(v,\lambda_0)]_G}{\partial \nu^n}\bigg|_{v=0},
\end{equation}
the following general relation holds
\begin{equation}
\frac{\partial^n \langle \hat{M}_z \rangle_G}{\partial \lambda_0^n}=\beta^n [C_{n+1}(\lambda_0)]_G.
\label{theoremcumulants}
\end{equation}
where $\langle M_z \rangle_G$ is the average magnetization of the system over the thermal state $\hat{\rho}_G(\lambda_0)$.
Thus, Eq.~\eqref{susceptibility} is exactly the $n=1$ case of Eq.~\eqref{theoremcumulants}. This relation is very important as it allows us to give a physical interpretation to the cumulants of the distribution for the system magnetization. To do this we can think of a magnetic material, e.g. a classical magnet, for which the commutation property holds, placed in a magnetic field $\lambda_0$, and we increase the magnetic field of the quantity $\Delta \lambda=\lambda_{\tau}-\lambda_0$. In this scenario the magnetization can be expressed in terms of a power series of the applied field variation as
\begin{equation}
\begin{aligned}
\langle \hat{M}_z \rangle_G(\lambda_{\tau})&=\langle \hat{M}_z \rangle_G(\lambda_0)+\chi_M^{(1)}(\lambda_0)\Delta \lambda+\chi_M^{(2)}(\lambda_0)\Delta \lambda^2+\\
&+\chi_M^{(3)}(\lambda_0)\Delta \lambda^3+...\\
\end{aligned}
\end{equation}
where
\begin{equation}
\label{chiJ}
\chi_M^{(j)}(\lambda_0)=\frac{1}{j!}\frac{\partial^j \langle \hat{M}_z \rangle_G}{\partial \lambda^j}\bigg|_{\lambda=\lambda_0}
\end{equation}
is the j-th order magnetic susceptibility at field $\lambda_0$. Comparing Eqs.~\eqref{theoremcumulants} and \eqref{chiJ} we get
\begin{equation}
\label{cumunuova}
[C_{n+1}(\lambda_0)]_G=\frac{n!}{\beta^n}\chi_M^{(n)}(\lambda_0).
\end{equation}
From linear response theory we know that the first order magnetic susceptibility $\chi_M^{(1)}(\lambda_0)$ is sufficient to characterize the response of the system to a small-amplitude external magnetic field around $\lambda_0$ \cite{reichl}. Here, differently, we are pushing the system far from equilibrium by applying a strong field variation $\Delta \lambda$, so that we need  the magnetic susceptibilities at every order to characterise the full response of the system. Now we can say that it is possible to interpret any cumulant of order $(n+1)$ of the magnetization distribution generated by the thermal state $\hat{\rho}_G(\lambda_0)$, as the $n^{\text{th}}$-order magnetic susceptibility of the system at the respective field $\lambda_0$.

Going back to the generic initial projected state $\hat{\rho}_0'$, a definition of the characteristic function in this case, equivalent to the one given for the thermal state in Eq. \eqref{caraG}, can be given simply as
\begin{equation}
G (v)= \tr{e^{i v \hat{M}_z} \hat{\rho}_0'}.
\label{caraGgeneric}
\end{equation}
Thus the moments $\langle \hat{M}_z^n \rangle$ and the cumulants $C_n$ of the magnetization distribution generated by a generic $\hat{\rho}_0'$ can be obtained by making use of this last function.
Keeping Eqs.~\eqref{firstwork} and \eqref{secondworkGeneric} in mind, it is easy to show that in the commuting case we have
\begin{equation}
\langle W^n \rangle=(-\Delta \lambda)^n \langle \hat{M}_z^n \rangle
\label{www}
\end{equation}
for any value of $n$ and any $\hat{\rho}_0'$. Furthermore, for the commuting case we have the equivalent relation for the cumulants
\begin{equation}
K_n = (-\Delta \lambda)^n C_n.
\label{ccc}
\end{equation}
Thus, if we are concerned with a thermal state, thanks to Eqs.~\eqref{cumunuova} and \eqref{ccc}, we can say that \emph{in the commuting case the $(n+1)^{\text th}$ cumulant of the work distribution is proportional to the $n^{\text th}$ order magnetic susceptibility}.\\
For a generic state, attaching a physical meaning to the cumulants of work is instead not as simple as the thermal case. However, for the commuting case, independently on the initial state, we have shown that the whole statistics of the work is entirely obtainable from the statistics of an observable, the magnetisation $\hat{M}_z$ over the initial projected state. So, in order to retrieve the statistics of work, we need first to project the initial state over the eigenbasis of the initial Hamiltonian, and then measure the magnetization.
So, for a generic initial state, \emph{it is not possible to obtain the statistics of work with just single projective measurements, even for the simple case of the sudden quench}. This result is non trivial as for the sudden quench we know that the state of the system freezes, so that we could have argued, intuitively, that single projective measurements should have been sufficient. For a thermal state instead, or in general for a state $\hat{\rho}_0$ such that $[\hat{\rho}_0,\hat{{\cal H}}(\lambda_0)]=0$, the projected state coincides with the actual initial one, so that in order to reconstruct the statistics of work we need to do just single projective measurements.

In the commuting case, the results for the time general dependent protocol are the same as for the sudden quench. In fact, for the time dependent protocol the results obtained so far are formally the same as long as we express the final Hamiltonian in the Heisenberg representation. However, such a change of picture is immaterial if the Hamiltonian  commutes with itself at every time.

\subsection{Non-Commuting case: $[\hat{\cal H}_{ss},\hat M_z]\ne0$}
In the case of $[\hat{\cal H}_{ss},\hat M_z]\ne0$, i.e. the case of a transverse magnetisation, Eq.~\eqref{www} and \eqref{ccc} do not hold anymore. In fact, it can be shown (cf. Appendix B) that the correct expression for the $n^{\text{th}}$ moment of the work distribution reads
\begin{equation}
\label{nthmoment}
\langle W^n \rangle =
\tr{\sum_{k=0}^n (-1)^k\binom{n}{k}\, \hat{\cal H}(\lambda_{\tau})^{(n-k)}\hat{\cal H}(\lambda_{0})^{k} \hat{\rho}_0'},~~~\forall n\in{\mathbb N}.
\end{equation}
We can see that $\langle W^n \rangle=\langle \bigl(\hat{\cal H}(\lambda_{\tau})-\hat{\cal H}(\lambda_{0})\bigr)^n \rangle=\left(-\Delta \lambda\right)^n \langle \hat{M}_z^n\rangle$ holds

\[ \text{for}~\left\{ 
  \begin{array}{l l}
    n=1,2 & \quad \text{if $[\hat{{\cal H}}(\lambda_{t}),\hat{\cal H}(\lambda_0)]\ne 0$}
   \\
    \forall n \in \mathbb{N} & \quad \text{if $[\hat{{\cal H}}(\lambda_{t}),\hat{\cal H}(\lambda_0)]= 0$~~.}
  \end{array} \right.\]
For a time dependent protocol Eq.~\eqref{nthmoment} still holds as long as the final Hamiltonian  $\hat{\cal H}(\lambda_{\tau})$ is replaced by its Heisenberg representation $\hat{\cal H}_H(\lambda_{\tau})=\hat{U}_\tau^{\dagger} \hat{H}(\lambda_{\tau}) \hat{U}_\tau$.
For a thermal state then, or in general for a state $\hat{\rho}_0$ such that $\hat{\rho}_0=\hat{\rho}_0'$, also in the non-commuting case the first two moments of the work are given by the average magnetization and the average of the square of the magnetization. However, this time we need to pay attention to the physical interpretation of these relations. In fact although it is still valid that 
\begin{equation}
\Delta W^2=\Delta \lambda^2  \Delta \hat{M}_z^2,
\label{variancevariance}
\end{equation}
the relation in Eq.~\eqref{susceptibility} does not hold anymore. In fact we show in Appendix C that the magnetic susceptibility, in the non-commuting case, can be written as 
\begin{equation}
\chi_M = \beta [\Delta \hat{M}_z^2]_G+\widetilde{\chi}_M,
\label{suscenuova}
\end{equation}
where the correction term $\widetilde{\chi}_M$ is given by \footnote{In the general case the magnetic susceptibility is given by the formula that can be found in R. Kubo, J. Phys. Soc. Jpn. \textbf{12},� 570--586� (1957)}
\begin{equation}
\begin{aligned}
&\widetilde{\chi}_M=\frac{1}{Z(\lambda_0)}\times\\
&{\tr{\sum_{n=1}^\infty \sum_{k=0}^{n-1} \frac{(-\beta)^n}{n!} \left[(\hat{H}_{ss}-\lambda_0 \hat{M}_z)^k,\hat{M}_z\right] \hat{M}_z \left(\hat{H}_{ss}-\lambda_0 \hat{M}_z\right)^{n-k-1}}}.
\end{aligned}
\end{equation}
On one hand, this shows that even the simple case of a sudden quench bears important consequences, as far as the statistics of work is involved. In fact, although it is a well-understood fact that work is not a quantum observable~\cite{lutz}, one could wonder whether specific protocols exist such that the full statistics of work could be reproduced by the statistics of a properly chosen quantum observable over the initial state, therefore enabling its direct assessment via single projective measurements. For example it is known that multiple-time probabilities can be recovered from a one-time probability of a larger system \cite{Rovelli}.
We showed that such a possibility is offered, for a sudden quench over a state which commutes with the initial Hamiltonian, by a magnetization which commutes with the interaction part of the model. However by looking at Eq.~\eqref{nthmoment} we see that, for the special case of $\hat{\rho}_0'=\hat{\rho}_0$ (e.g. the initial thermal state), for every moment of the work we need to do single projective measurements of an observable, which however in the non-commuting case has not such a simple physical meaning as the magnetization for the commuting case. 

The idea of going beyond the two-time measurement approach is motivated by the issues concerning the applicability of fluctuation theorems when some of the typical assumptions made (e.g. initial thermal states, closed system dynamics) are relaxed. Generalizations of fluctuation theorems along the lines of non-commuting states and observables have been put forward in by Kafri and Deffner in Ref.~\cite{varie}. In Ref.~\cite{watanabe} Watanabe {\it et al.} show that, by using a particular type of generalized energy measurements, the resulting work statistics is simply related to that of projective measurements. Recently conditions have been given also for the fluctuating work to be physically meaningful for a system that starts its evolution from a non-equilibrium state \cite{allah2}. A different approach instead deals with the formulation of new fluctuation theorems when the system is not described by a (micro) canonical density matrix but it is described by a (micro) canonical distribution of wave functions \cite{campisinjp}. A fluctuation theorem for the nonequilibrium entropy production in quantum phase space is instead derived in Ref.~\cite{deffnerQPS}, which enables a thermodynamic description of open and closed quantum systems. Also several works recently focused on the thermodynamic description of fully open quantum systems by making use of the quantum jump, quantum trajectory description of the evolution of the system \cite{leggio,Horowitz,Hekking}.\\

As a last remark of the Section we stress that the differences between the commuting and non-commuting cases can be seen also as a direct application to the case of a many-body system of the comparison of two different definitions of work given in Refs.~\cite{Allahverdyan,nolte}. The first definition is the most common one given in Eq.~\eqref{defwork}. In this case we know that the work is a stochastic variable. The second definition deals instead with a work operator $\Delta \hat{E}(\lambda_0,\lambda_{\tau})=\hat{U}_\tau^{\dagger}\hat{\cal H}(\lambda_{\tau})\hat{U}_\tau-\hat{\cal H}(\lambda_0)$, but the Jarzynski equation is known to hold only for the first definition of work. Indeed, while the average of any linear and quadratic function of the two definitions of works are the same, higher order functions reproduce the same average only in the case in which the initial and final Hamiltonians commute. Thus, for the many-body model $\hat{\cal H}(\lambda_t)=\hat{\cal H}_{ss}-\lambda_t \hat{M}_z$, the work operator is given by
\begin{equation}
\Delta \hat{E}(\lambda_0,\lambda_{\tau})=\hat{U}_\tau^{\dagger}\hat{\cal H}_{ss}\hat{U}_\tau-\hat{\cal H}_{ss}-\lambda_{\tau}\hat{U}_\tau^{\dagger}\hat{M}_{z}\hat{U}_\tau+\lambda_0 \hat{M}_{z}.
\end{equation}
For the sudden quench we have
\begin{equation}
\Delta \hat{E}(\lambda_0,\lambda_{\tau})=-\Delta \lambda \hat{M}_z
\end{equation}
so that if in line with Eq.~\eqref{caraG} we define the characteristic function of the work operator for the sudden quench
\begin{equation}
\chi_{\Delta \hat{E}} (v,\lambda_0)= \frac{\tr{e^{i v \Delta \hat{E}} e^{-\beta(\hat{\mathcal H}_{ss}-\lambda_0 \hat{M}_z)}}}{\mathcal{Z}(\lambda_0)}
\label{caraDelta}
\end{equation}
and the associated cumulants $(K_{\Delta \hat{E}})_n=i^{-n}\partial^n_\nu\log \chi_{\Delta \hat{E}}|_{v=0}$, Eq.~\eqref{theoremcumulants} can be interpreted as a statement on the cumulants of the characteristic function of the work operator for the sudden quench
\begin{equation}
\partial_{\lambda_0}^n \langle \Delta \hat{E} \rangle=\beta^n (-\Delta \lambda)^{-n} (K_{\Delta \hat{E}})_{n+1}.
\label{theoremcumulantsDelta}
\end{equation}

\subsection{Assessing the non-equilibrium thermodynamics via the cumulants of the work distribution}
We have clearified the useful role played by the initial thermal state over the possibility of getting the statistics of work from single projective measurements, and the physical meaning of the cumulants of the work done on a system whose initial state is thermal. Keeping in mind these results we now focus just on this state and so we will not use anymore the subscript $G$ to indicate averages over thermal states.
The importance of looking at the full statistics of the work distribution is clear from the point of view of non-equilibrium thermodynamics. In fact we can use the Jarzynski equality in the form $\Delta F=-(1/\beta)\log\langle e^{-\beta W} \rangle$ to show that we can express the free-energy difference $\Delta F$ in term of a sum of cumulants $K_n$ of the work distribution as \cite{Jarzynski}
\begin{equation}
\Delta F=\sum_{n=1}^{\infty}\frac{(-\beta)^{n-1}}{n!}K_n(\beta).
\label{deltaf}
\end{equation}
The non-equilibrium nature of the transformation that we are addressing here allows us to recast the second principle of thermodynamics as $\langle W \rangle \ge \Delta F$, which suggests the existence of an {\it irreversible} form of work defined as $\langle W_{\text{diss}} \rangle = \langle W \rangle -\Delta F$. In turn, this allows the introduction of the ``non-equilibrium lag'' $L_{\text{irr}}=\beta \langle W_{\text{diss}}\rangle=\beta (\langle W \rangle -\Delta F)$ that quantifies the degree of irreversibility of the quenched dynamics in terms of the actual state lag between the actual state $\hat{\rho}_t$ of the system at a given time of the evolution and the hypothetical thermal equilibrium state $\hat{\rho}_t^{\text{eq}}$ associated with the Hamiltonian of the system at that time. In fact, it can be shown that $L_{\text{irr}}=\Delta S[\hat{\rho}_t||\hat{\rho}_t^{eq}]$ with $\Delta S[\hat{\rho} ||\hat{\sigma}]=\text{Tr}(\hat{\rho}\log\hat{\rho}-\hat{\rho}\log\hat{\sigma})$ the relative entropy between two arbitrary states $\hat{\rho}$ and $\hat{\sigma}$~\cite{Bochkov81aPHYSA106,Schloegl66ZP191,Vaikuntanathan09EPL87,Deffner10PRL105}. The non-equilibrium lag can be cast in terms of the set of cumulants of $\{K_{n\ge2}\}$ as
\begin{equation}
L_{\text{irr}}=\sum_{n=2}^{\infty} \frac{(-\beta)^n}{n!}K_n(\beta).
\label{deltaentropy}
\end{equation}
In the expressions above, we have explicitly shown the dependence of the cumulants on the inverse temperature $\beta$ in order to stress that these formulas do not have the form of a power series expansion with respect to $\beta$. It is in fact clear in Eq.~\eqref{cumulants} that the cumulants depend on $\beta$ via the characteristic function. Eqs.~\eqref{deltaf} and \eqref{deltaentropy}  allow us to see clearly how the cumulants of the work distribution are related to the free-energy and the non-equilibrium lag.
Notably, the expression we obtained above is a generalisation of the Eq.~$(60)$ in Ref.~\cite{Sotiriadis}, which was found by the authors just for the case of a small quench. Indeed if we take just the first term of the expression above, relating the variance of the work to the derivatives of the free-energy, we obtain the very same expression. The expression we obtained above is instead valid for every \emph{size} of the quench.

To fix the ideas, let us assume that the spin-spin part of the Hamiltonian in Eq.~\eqref{egmodel} commutes with the term $\lambda_t\sum^N_{i=1}\hat \sigma^z_i$, which thus embodies the longitudinal magnetization of the system. Referring to Eqs.~\eqref{deltaf} and \eqref{deltaentropy}, we can now say that, in this case, the whole non-equilibrium thermodynamics of the system can be obtained from the full statistics of the magnetization itself and so via single projective measurements. Being in the commuting case, we can refer to the time dependent protocol and not just the sudden quench. 
This highlights in a physically very clear way the qualitative difference arising from considering quenched operators that do commute with the unperturbed Hamiltonian. Moreover, we can consider  the possibility of finding signatures of critical behaviour of a quantum many-body model by investigating the statistics of a thermodynamical quantity. We know a priori that the statistics of work following a global quench of the longitudinal magnetization will indeed show evidence of critical behaviour. In fact, the statistics of work coincides with that of the order parameter. In particular if we are dealing with an $n^{\text{th}}$ order phase transition, the $(n-1)^{\text{th}}$ derivative of the order parameter will be discontinuous.

We showed that in the commuting case the thermodynamics can be retrieved by looking just at the statistics of a single quantum observable. This happens typically for first order quantum phase transitions that occur for Hamiltonians that are the sum of two competing and commuting terms giving rise, for any system size, to energy crossings. Second order phase transitions instead emerge, for increasing number of particles, from the competition between two non commuting operators. Crucially in this case the statistics of work beyond its second moment cannot be interpreted in terms of a simple quantum observable (cf. Eq.~\eqref{nthmoment}). This ultimately can be ascribed to the intrinsic non-commutativity of quantum mechanics. 
In fact for instantaneous quenches in classical systems the statistics of work can always be mapped onto the equilibrium fluctuations of a classical observable, namely the difference of post and pre-quench Hamiltonians $\Delta E(\boldsymbol{z})={\cal H}_1(\boldsymbol{z})-{\cal H}_0(\boldsymbol{z})$, where $\boldsymbol{z}$ is a point in the phase space of the classical system
\footnote{For a classical instantaneous quench ${\cal H}_0(\boldsymbol{z}) \rightarrow {\cal H}_1(\boldsymbol{z})={\cal H}_0(\boldsymbol{z})+\Delta E(\boldsymbol{z})$, the classical work generating function reads \cite{campisi}: $G_{cl}(u)=\int d\boldsymbol{z} \exp\left[i u \left( {\cal H}_1(\boldsymbol{z})-{\cal H}_0(\boldsymbol{z}) \right)\right]\exp\left[-\beta {\cal H}_0(\boldsymbol{z})\right]/{\cal Z}=\int d\boldsymbol{z} \exp\left[ i u \Delta E(\boldsymbol z) \right]\exp\left[ -\beta {\cal H}_0(\boldsymbol{z})/{\cal Z} \right]$ which is the generating function of the statistics of the classical observable $\Delta E(\boldsymbol{z})$.}.
This feature can be used as a witness of quantumness in the system. For example, if the statistics of the observable $\Delta E(\boldsymbol{z})$ does not obey the Jarzynski identity $\langle e^{-\beta \Delta E} \rangle = e^{-\beta \Delta F}$, then the system is non-classical. 
As an example of evaluation of the probability distribution of an Ising-like system with mean-field interaction can be found in Ref. \cite{Imparato}. In the case of the Ising model in a transverse field, which will be introduced later in this paper, the classical counterpart of the model has qualitative differences with the quantum case which go beyond the difference just in the commutation between operators. An indirect confirmation of this can be seen in the fact that the classical model gives a magnetization which is different from the quantum case \cite{verrucchi}. 
In light of our results, should the only difference between a given classical model and its quantum counterpart be in the commutation relations between the respective operators, the first two moments of the corresponding quantum and classical distributions of work should be equal.

\begin{figure}[t]
\includegraphics[width=8cm]{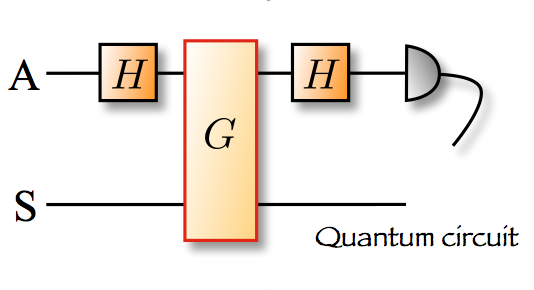}
\caption{Circuit representation of the interferometric scheme used to reconstruct the characteristic function of the magnetisation distribution. H represents a Hadamard gate, while G is the main gate in the circuit given in Eq.~\eqref{gate}.} 
\label{circuito} 
\end{figure}

As a final remark we want to let the reader notice that the characteristic function of the distribution generated by the magnetization, defined in Eq. \eqref{caraG}, can be reconstructed also using the very same setup suggested in Ref.~\cite{Oxf} for the measurement of the characteristic function of work. The scheme is reported in Fig. \ref{circuito} and implies the interaction, through a conditional gate $\hat G(u)$, of a suitably prepared controllable ancilla $A$ with the  system $S$ under scrutiny.
 This is indeed exactly the case of the simple illustrative case analysed by Mazzola {\it et al.} in~\cite{Oxf}, where the initial and final Hamiltonians commute, so that the gate $G$ in this case is simply given by
\begin{equation}
\hat{G}(u)=\mathbb{1} \otimes | 0 \rangle \langle 0 |_A + e^{-i M_z u} \otimes | 1 \rangle \langle 1 |_A
\label{gate}
\end{equation}
and $\{| 0 \rangle_A,| 1 \rangle_A  \}$ is a basis of egeinstates of the Hilbert space of the ancillary system used in the scheme. The implementation of this proposal when $S$ is embodied by a quantum many-body system requires some considerations. In fact, it would be particularly convenient to let the ancilla interact only with one element of the many-body system, so as to reduce the complexity of the implementation and the control required. A possible way forward is based on the adaptation of the proposal discussed in Ref. \cite{Kavan}. A detailed description of the protocol is beyond the scopes of this investigation and will be reported elsewhere.

\section{The transverse Ising model as a special case study}
\label{model}
 
Having identified clearly the conditions under which the statistics of work could reveal many-body features, in this Section we provide a clear example of a model for which an observable that indeed does not commute with the spin-spin Hamiltonian term remains, nevertheless, quite informative. We have shown in Sec. III that the second cumulant of the work, for the non-commuting case, is not simply proportional to the magnetic susceptibility of the model (cf. Eq.~\eqref{suscenuova}). However the second cumulant of the work is proportional to the second cumulant of the magnetization (cf. Eq.~\eqref{variancevariance}). If we instead move already to the third moment (cumulant) of the work, this is not even proportional to the third moment (cumulant) of the magnetization (cf. Eq.~\eqref{nthmoment}). This simple, yet fundamental, observation makes the question whether we can observe signatures of quantum criticality in higher order cumulants meaningful and worth analyzing and in particular this is one of the most interesting and direct applications of our results found in the previous Sections.

We assess the statistics of work in a quantum Ising model initially prepared in a thermal state and subjected to a sudden quench of a {\it transverse} magnetic field. More specifically, we consider a one-dimensional ring of $N$ spin-$1/2$ particles that interact with their nearest neighbours via a ferromagnetic coupling along the $x$ axis and with an external field applied along the $z$ axis. The zero-temperature version of such paradigmatic situation has been examined in Ref.~\cite{Silva}, while Dorner {\it et al.} \cite{Dorner} have analysed the non-zero temperature case to get insight into the dissipated work. The Hamiltonian model reads
\begin{equation}
\hat{{\cal H}}(\lambda)=-\sum_{i=1}^N \hat{\sigma}_i^x\hat{\sigma}_{i+1}^x -\lambda \sum_{i=1}^N \hat{\sigma}_i^z,
\label{ising}
\end{equation}
where $\hat\sigma^{k}_{N+1}=\hat\sigma^k_1~(k=x,y,z)$. In the thermodynamic limit $N\to\infty$ and at $T$=0, the spin system undergoes a second-order phase transition at the critical value $\lambda_c=1$. The critical point separates a ferromagnetic phase at $\lambda<1$, where the ground state is doubly degenerate (the spins all point in either the positive or negative $x$ direction), from a paramagnetic phase at $\lambda>1$ with a non-degenerate ground state characterised by all the spins aligned with the magnetic field.

Following the formalism introduced in Ref.~\cite{Dorner}, we report in Appendix D the typical diagonalisation procedure for the Ising model in a transverse field. The diagonal form of the \emph{pre-quench} Hamiltonian reads
\begin{equation}
\hat{\cal H}(\lambda_0)=\sum_{k \in K^+}\epsilon_k(\lambda_0) \biggl( \hat{\gamma}_k^{\dagger}\hat{\gamma}_k-\frac{1}{2} \biggr)
\label{diagonalHpre}
\end{equation}
with $K^+=\{k=\pm\pi(2n-1)/N\}$ and $n=1,...,{N}/{2}$, as we are restricting our attention to the even parity subspace of the model, and $\hat{\gamma}_k,\hat{\gamma}_k^{\dagger}$ are fermionic operators labelled by the values of pseudomomenta in the set $K^+$. The \emph{post-quench} Hamiltonian is found to be given by the diagonal model
\begin{equation}
\hat{\cal H}(\lambda_{\tau})=\sum_{k \in K^+}\epsilon_k(\lambda_{\tau}) \biggl( \hat{\gamma}_k'^{\dagger}\hat{\gamma}'_k-\frac{1}{2} \biggr)
\label{diagonalHpost}
\end{equation}
where the fermionic operators $\hat{\gamma}_k',\hat{\gamma}_k'^{\dagger}$ are different from their \emph{pre-quench counterpart} $\hat{\gamma}_k, \hat{\gamma}_k^{\dagger}$.
The characteristic function for this system is obtained by evaluating the trace in Eq.~\eqref{characteristicfunctionsudden} over the eigenstates of the initial Hamiltonian with the result 
\begin{equation}
\begin{aligned}
\chi(u, \tau)&=\frac{1}{{\cal Z}(\lambda_0)}\prod_{\substack{k \in K^+\\k>0}} \Biggl\{ e^{(iu+\beta)\epsilon_k (\lambda_0)}\biggl[ C_k^- (u,\lambda_{\tau})+S_k^+ (u,\lambda_{\tau}) \biggr]+\\
&+e^{-(iu+\beta)\epsilon_k (\lambda_0)}\biggl[ C_k^+ (u,\lambda_{\tau})+S_k^- (u,\lambda_{\tau}) \biggr]+2 \Biggr\}
\label{IsingChi}
\end{aligned}
\end{equation}
where $C_k^{\pm}=\cos^2(\Delta_k/2)e^{\pm iu\epsilon_k(\lambda)}$, $S_k^{\pm}=\sin^2(\Delta_k/2)e^{\pm iu\epsilon_k(\lambda)}$ and $\Delta_k=\phi_k'-\phi_k$ is the difference in the pre- and post-quench Bogoliubov angles. The availability of the analytical expression of the characteristic function allows for the exact evaluation of both the cumulants and the moments of the work distribution.

In what follows we focus on the occurrence of signatures of a quantum phase transition in the statistics of the work done on the system by means of the quenched process. Although quantum phase transitions are rigorously defined by the non analyticity of the energy of the ground state with respect to a Hamiltonian parameter, evidence of their occurrence at finite temperature can be found ~\cite{sachdev}. It is in this spirit that we will develop our analysis, i.e. by first studying the case of $T=0$ and then moving towards a non-zero temperature scenario to see the emergence of irreversibility from the microscopic quantum fluctuations responsible for the occurrence of the quantum phase transition. 
\begin{figure}[t]
\includegraphics[width=8cm]{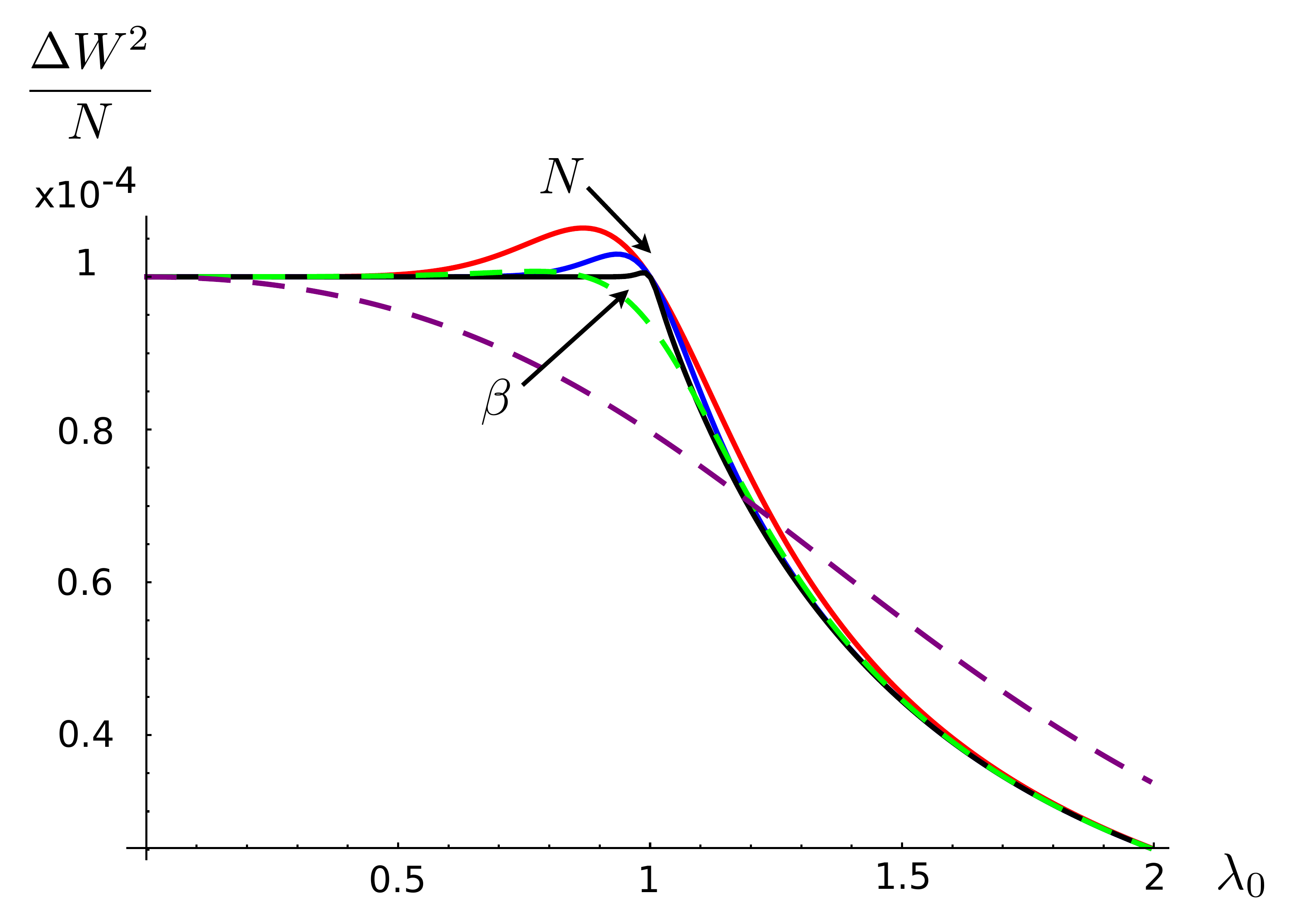}
\caption{(Color online) Normalised variance of the work distribution $\Delta W^2/N$ plotted against the initial value of the magnetic field $\lambda_0$. The solid lines are for different number of spins (Red $N=10$, Blue $N=20$ and Black $N=100$) at inverse temperature $\beta=100$. The dashed lines are instead for different inverse temperatures (Purple $\beta=1$, Green $\beta=5$) for $N=100$ spins. The process consists of a sudden quench of amplitude $\Delta \lambda=0.01$, which is smaller than the minimum value of the gap in the model $\Delta E_{min}\approx 0.06$. All the energies are considered to be in units of the interaction coupling $J$ between the spins, since the model in Eq. \eqref{ising} is obtained indeed dividing the complete Hamiltonian by $J$, so that $\lambda$ is actually the ratio between the magnetic field and $J$.} 
\label{variaNT} 
\end{figure}

We start our investigation by studying the variance of the work distribution versus the initial magnetic field $\lambda_0$ [cf. Fig.~\ref{variaNT}]. A sharp transition from a flat region in the ferromagnetic phase to a monotonically decreasing region in the paramagnetic phase can be clearly seen. Furthermore, the transition becomes sharper as the size of the system grows. This is indicated in the figure by the $N$ arrow. This is due to the fact that, in the thermodynamic limit, the energy gap between the two lowest-lying states closes at the critical point. Correspondingly, a ferromagnetic phase transition is enforced. As the variance of the work distribution is proportional to the variance of the magnetization over the initial thermal state (cf. Eq.~\eqref{variancevariance}), such transition appears neatly in the behaviour of $\Delta \hat M^2_z/N$. We want to stress that this transition is not solely ascribed to the discontinuity of the transversal susceptibility of the model since, as we stressed earlier, the susceptibility of the model has an additional term $\widetilde{\chi}_M$, other than the variance of the magnetization (cf. Eq.~\eqref{suscenuova}).
In Fig.~\ref{variaNT} we also examine the influence of temperature on the variance: needless to say, albeit the same trend exhibited can still be appreciated, high temperatures clearly smoothen out the sharp edge of the transition, yet leaving it clearly recognizable. This is indicated by the $\beta$ arrow. 

\begin{figure}[t]
\includegraphics[width=8cm]{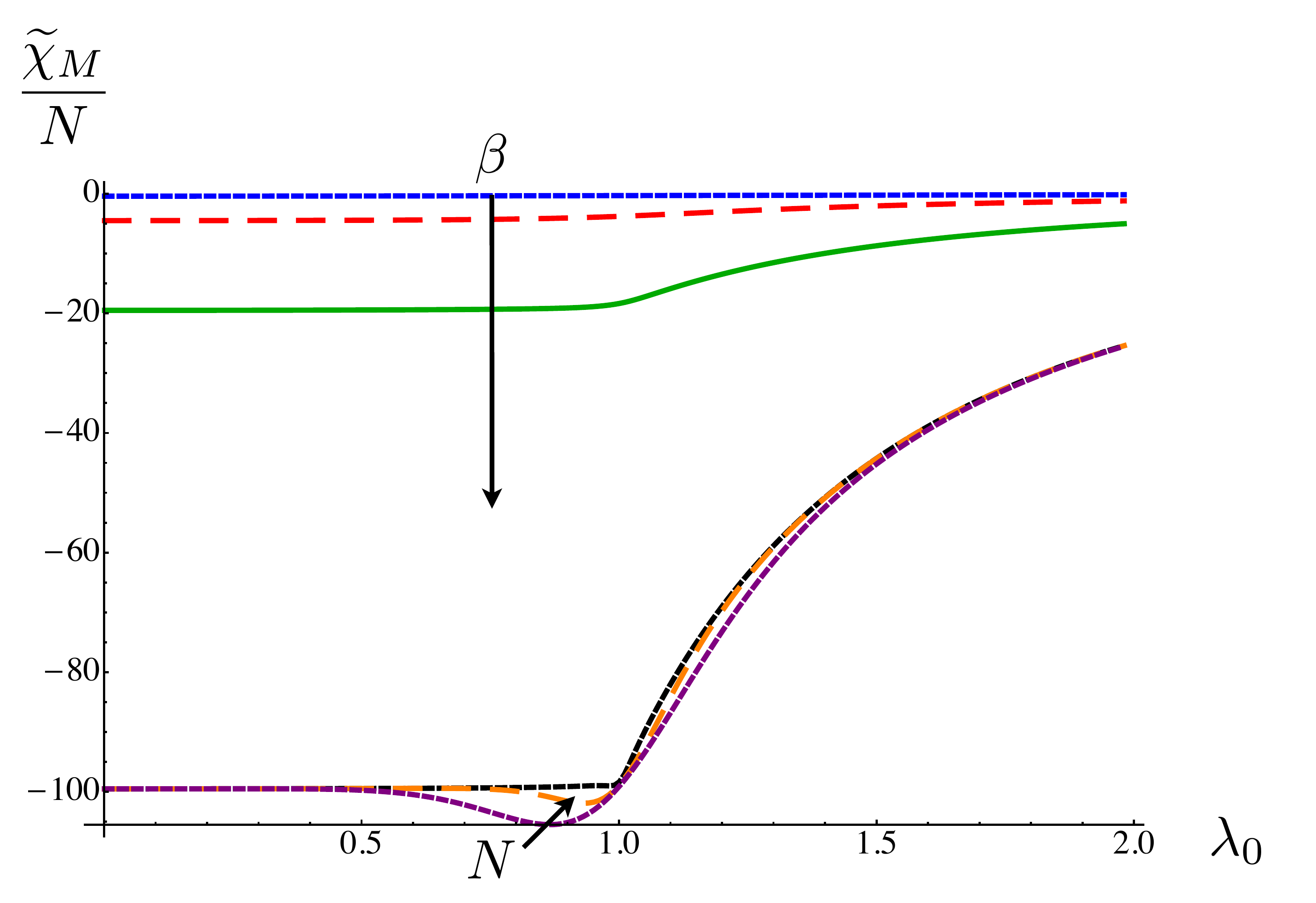}
\caption{(Color online) Additional term of the susceptibility $\widetilde{\chi}_M$, normalised with respect to the number of particles, arising from non the non-commutativity between the magnetization and the interaction part of the Hamiltonian. (Low part of the figure) The short dashed line (purple) is for $N=10$ and $\beta=100$, the long dashed line (orange) is for $N=20$ and $\beta=100$, the solid line (black) is for $N=100$ and $\beta=100$. (Upper part of the figure) The short dashed line (blue) is for $N=100$ and $\beta=1$, the long dashed line (red) is for $N=100$ and $\beta=5$, the solid line (green) is for $N=100$ and $\beta=20$. The quench is $\Delta \lambda=0.01$.} 
\label{chinuova} 
\end{figure}

In Fig.~\eqref{chinuova} we show the behaviour of the correction term $\widetilde\chi_M$, normalised with respect to the number of particles, arising from the non-commutativity between the magnetization and the interaction part of the Hamiltonian (cf. Eq.~\eqref{suscenuova}). We study this quantity with respect to the initial magnetic field $\lambda_0$, observing a transition-like behaviour when the temperature is lowered and the number of particles is increased. Notably $\widetilde{\chi}_M$ goes to zero as soon as the temperature is increased. Thus in the high temperature regime the system behaves classically and the correction term $\widetilde{\chi}_M$ goes to zero. Despite the term $\widetilde{\chi}_M$ being negative, the susceptibility $\chi_M$ is always positive and it shows the typical divergence behaviour near the critical point, that we do not report here.

\begin{figure}[t]
\includegraphics[width=8cm]{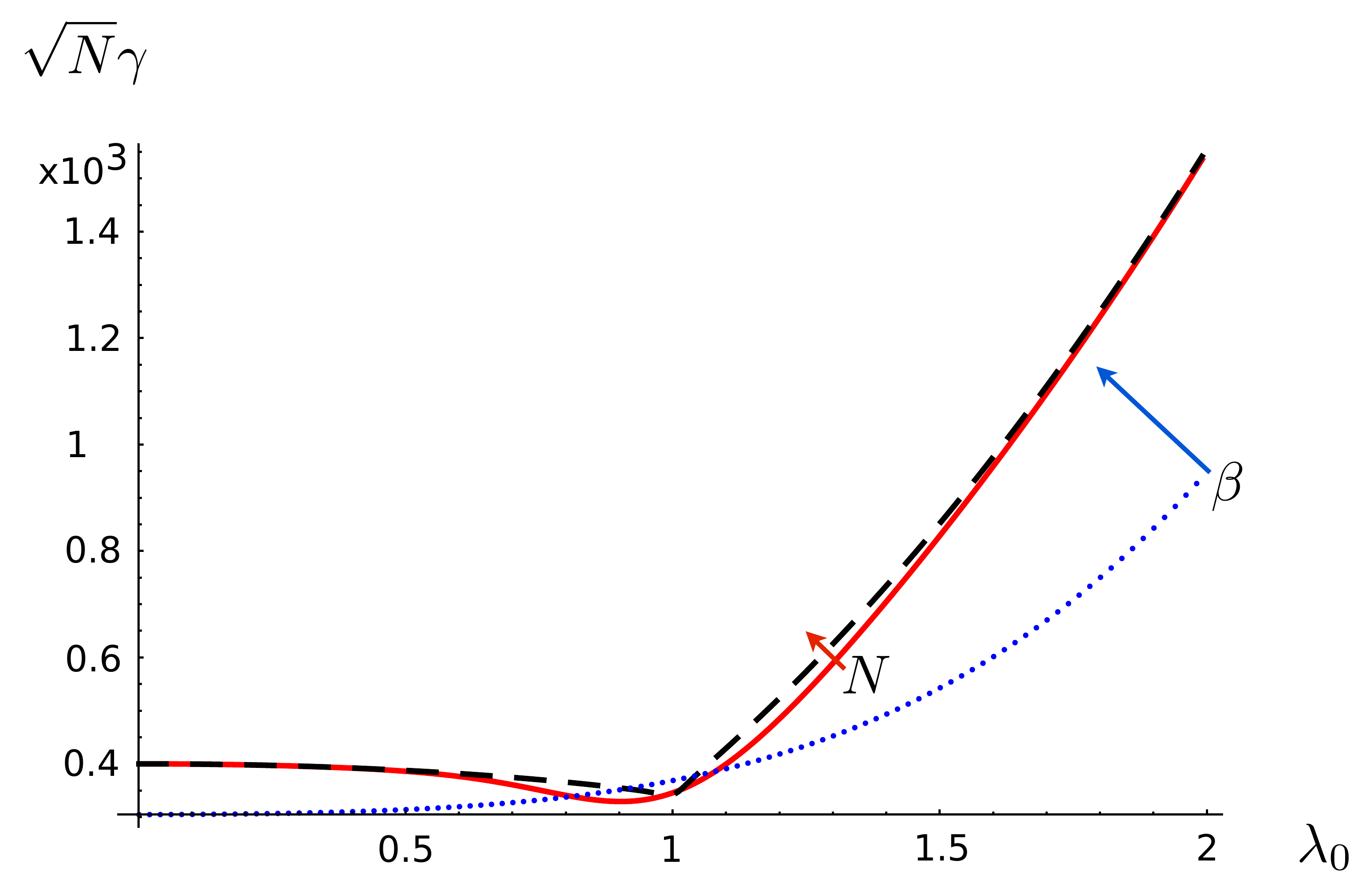}
\caption{(Color online) Normalised skewness of the work distribution $\gamma \sqrt{N}$ plotted against the initial value of the magnetic field $\lambda_0$. The dashed line is for $N=100$ and $\beta=100$,  the solid one is for $N=10$ and $\beta=100$, while the dotted one is for $N=100$ and $\beta=1$. The quench is $\Delta \lambda=0.01$.} 
\label{skewNT} 
\end{figure}

Fig.~\ref{skewNT} reports the trend followed by the normalised skewness $\gamma\sqrt{N}$ of the work distribution as the amplitude of the initial magnetic field $\lambda_0$ grows. Here $\gamma={K_3}/{\sigma^3}$ with $\sigma=\sqrt{\Delta \hat{M}_z^2}$ the standard deviation of the distribution. The skewness quantifies the asymmetry of a probability distribution, and is thus quite informative.
As $\gamma$ is always positive, we can infer that the area underneath the right tail of the distribution is larger than the one under the left one. We can see that the skewness, alongside the variance, has a transition from an almost flat region in the ferromagnetic phase to a monotonically increasing region in the paramagnetic phase. This qualitatevely means that the work distribution for a quench done on a spin chain initially in the paramagnetic phase is more asymmetric than if the system had been initially in the ferromagnetic phase. The transition is also sharper for higher number of spins (as indicated in Fig.~\ref{skewNT} by an arrow), again in light of the effects arising when approaching the thermodynamic limit.

Fig.~\ref{skewNT} reports also the skewness at a given number of spins for two different temperatures. The same behaviour can be appreciated, and considerations in line with those made for Fig.~\ref{variaNT} can be made. By varying the amplitude of the quench $\Delta \lambda$ we get qualitatevily similar results to those that we have already described, thus hinting at the independence of the results discussed so far from the \emph{size} of the quench. Thus, quite remarkably, we have demonstrated the possibility to use the variance and the skewness of the work distribution as witnesses of the quantum phase transition in a transverse Ising model. This is even more interesting in light of the fact that, as we showed, the variance of the work is not directly proportional to the susceptibility as in the commuting case, and the third cumulant entering the skewness is not even proportional to a power of magnetization (cf. Appendix B). Yet, the signatures of the quantum phase transitions are very evident.

\begin{figure}[h]
\includegraphics[width=8cm]{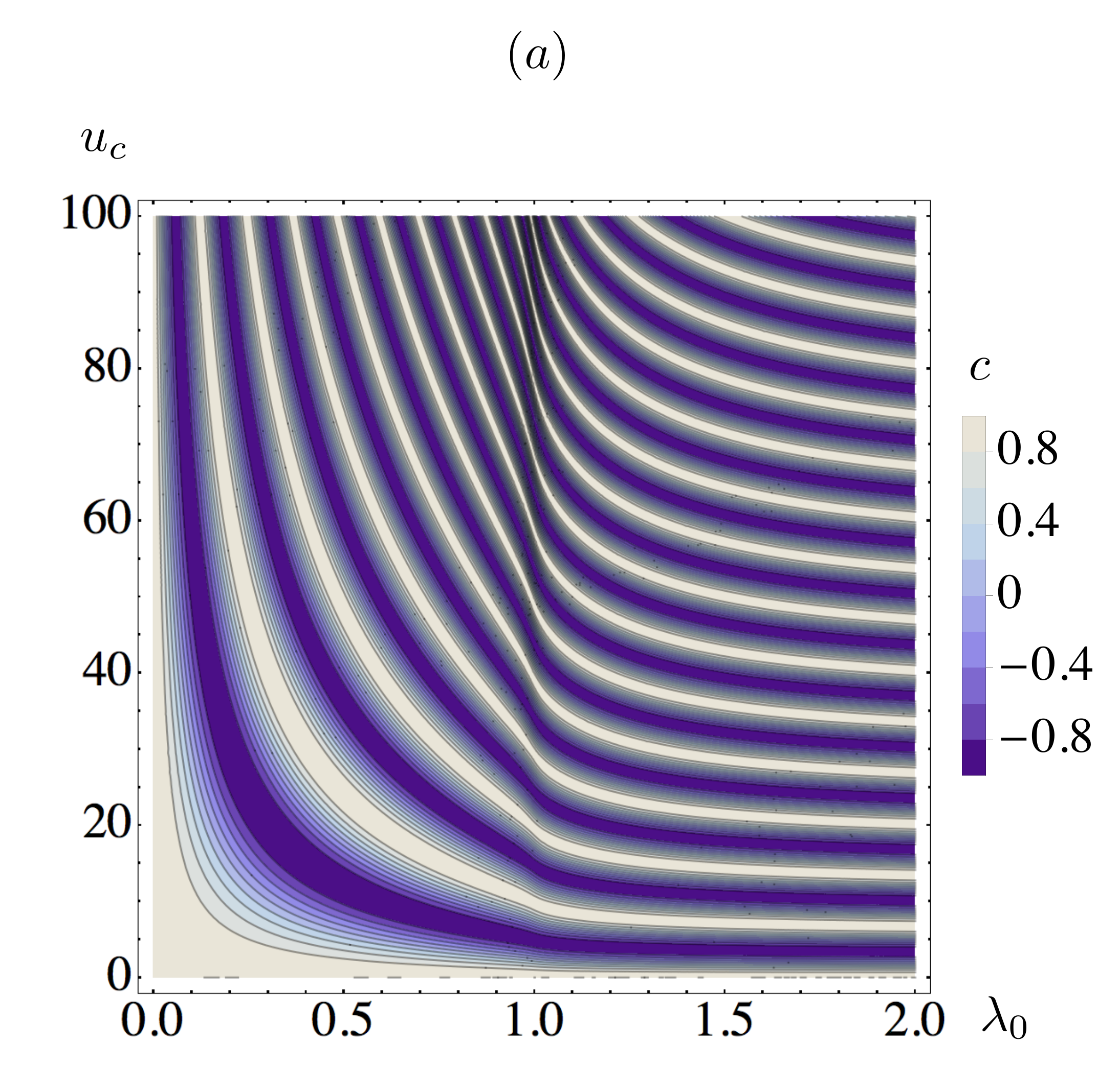}
\includegraphics[width=8cm]{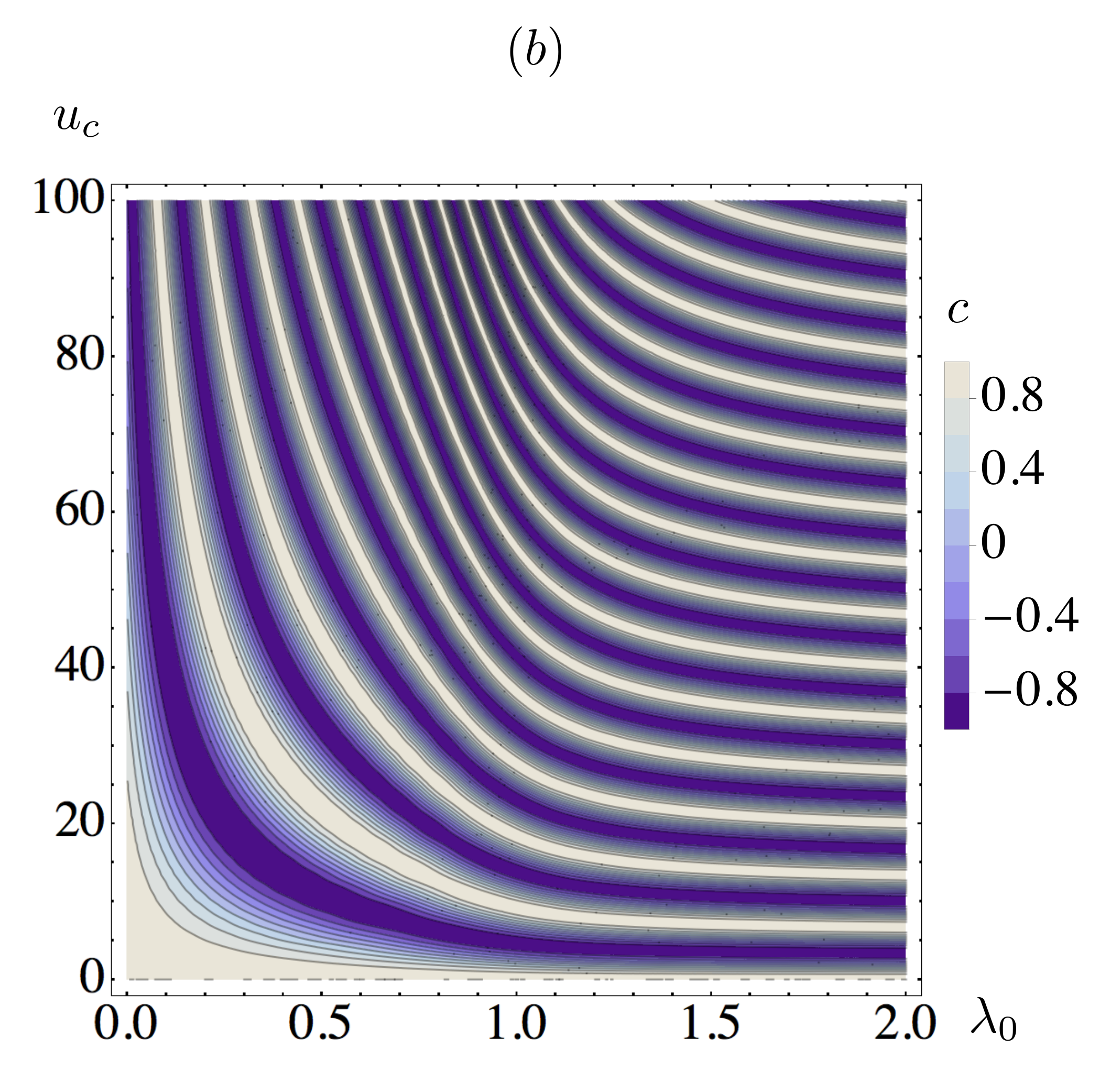}
\caption{(Color online) Contour plot of the real part of the characteristic function of work, i.e. plot of several equipotential curves (curves $u_c(\lambda_0)$ for which $\Re\{\chi(u_c,\lambda_0+\Delta\lambda)\}=c$). The curves are plotted for $\beta = 100$, various sizes of the system and amplitude of the quench. In panel {\it (a)} we take $N= 100, \Delta \lambda =0.01$, while in {\it (b)} we have $N=10, \Delta \lambda=0.1$.} 
\label{RealChiN} 
\end{figure}

Qualitatively very similar results have been found for any other higher cumulant of the work distribution that we have been able to address. The irreversible work was instead already studied in Ref.~\cite{Dorner}. As we did for the moments of the work, the authors analysed the scaling of the irreversible work with respect to the temperature and the size of the system, finding again similar results such as the presence of a marked signature of phase transition at the critical point of the thermodynamical limit of the model; a peak in that case. Also they found a difference of the value of the irreversible work between the two phases.\\
We now move to the assessment of the distribution itself. In Fig.~\ref{RealChiN} we show the contour plot of the real part of the characteristic function $\text{Re}[\chi(u)]$ for different sizes of the system. The contour plot is the plot of several equipotential curves, i.e. curves $u_c(\lambda_0)$ for which $\Re\{\chi(u_c,\lambda_0+\Delta\lambda)\}=c$). As $\text{Re}[\chi(u)]$ turns out to be an even function of $u$, we restrict our attention to positive values of this quantity.

A rather distinct functional behaviour of the characteristic function emerges between the ferromagnetic region ($\lambda_0 < 1$) and the paramagnetic one ($\lambda_0 >1$), with a discontinuity located approximately at the interface between the two regions. Although we have been able to study explicitly only the case of a finite number of spins (due to the difficulty inherent in the explicit evaluation of $\text{Re}[\chi(u)]$ for very large sizes of the system), the trend shown in Fig.~\ref{RealChiN} suggests a non-analytic behavior of the characteristic function at the critical point of the infinite model. In order to gather further evidence of such conjecture, we have turned to the numerical study of the derivative of the characteristic function with respect to the work parameter.

\begin{figure}[t]
\includegraphics[width=8cm]{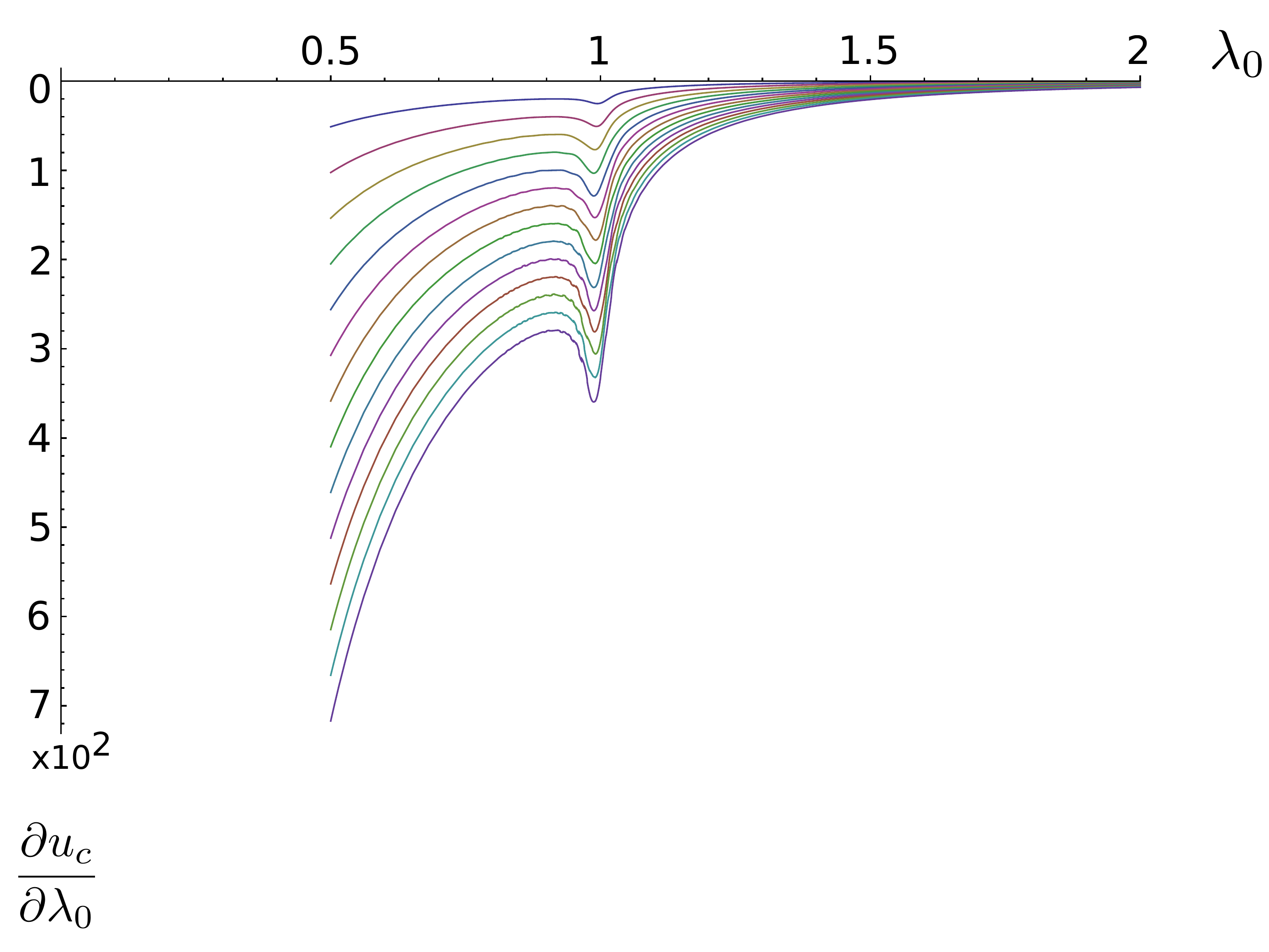} 
\caption{(Color online) First derivatives with respect to $\lambda_0$ of some equipotential curves of the real part of the characteristic functions for $N=100, \Delta \lambda=0.01$ (cf. Fig.~\ref{RealChiN} (a)). The equipotential curves are the curves $u_c(\lambda_0)$ for which $\Re\{\chi(u_c(\lambda_0),\lambda_0+\Delta\lambda)\}=c$.} 
\label{wit} 
\end{figure}

\begin{figure}[t]
\includegraphics[width=8cm]{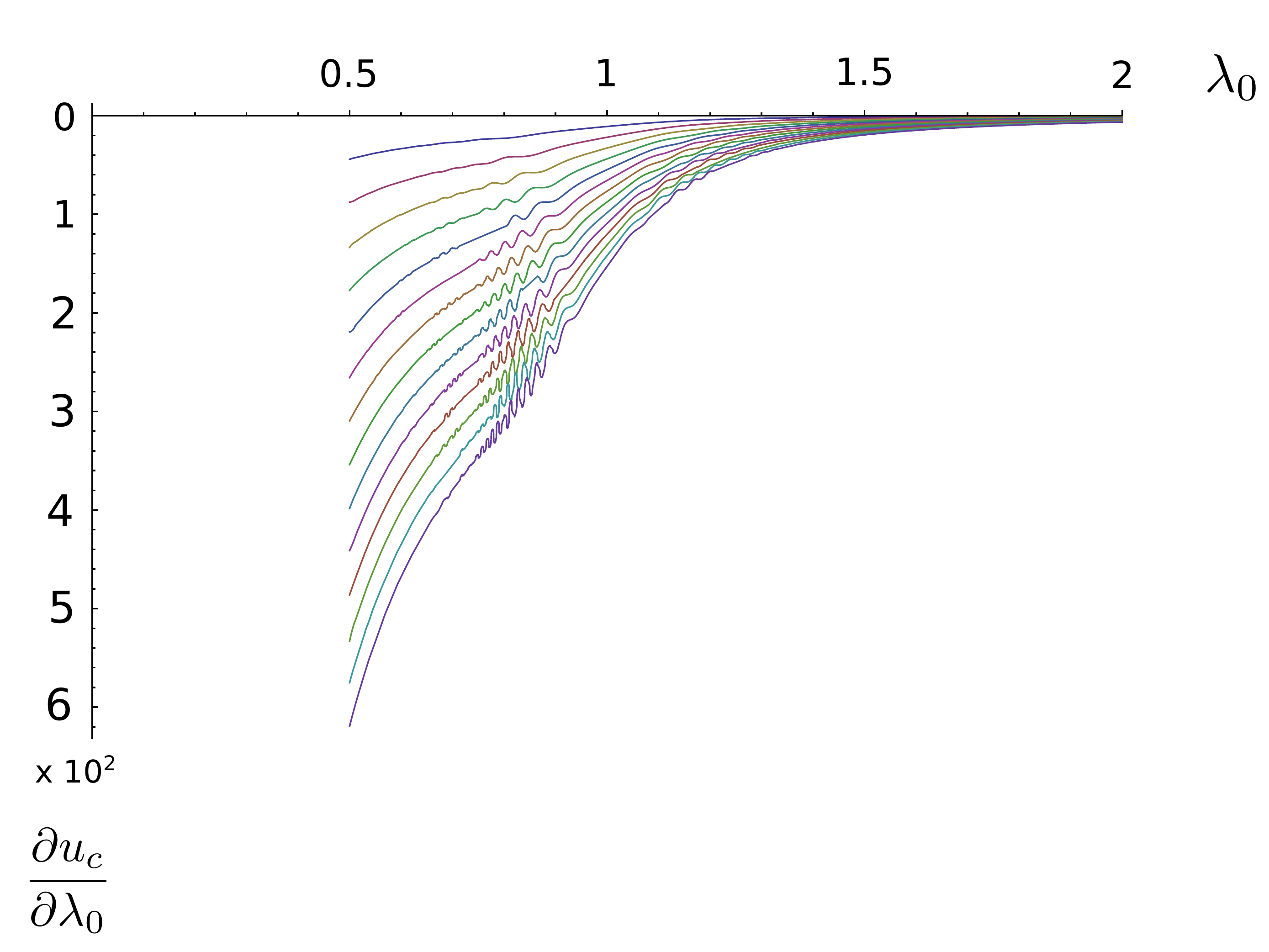} 
\caption{(Color online) First derivatives with respect to $\lambda_0$ of some equipotential curves of the real part of the characteristic functions for $N=10, \Delta \lambda=0.1$ (cf. Fig.~\ref{RealChiN} (b)). The equipotential curves are the curves $u_c(\lambda_0)$ for which $\Re\{\chi(u_c(\lambda_0),\lambda_0+\Delta\lambda)\}=c$.} 
\label{wit2} 
\end{figure}

In Fig.~\ref{wit} and Fig.~\ref{wit2} we show the derivatives of the curves extracted from the contour plots. We can see in Fig.~\ref{wit} that for $N=100$ the derivatives display a very pronounced change of behaviour in proximity of the critical point of the model, while in Fig.~\ref{wit2}, for the case $N=10$, it is shown quite evidently that this flattens out. Again, the change in behaviour is more neatly pronounced when the number of particles in the system grows and it gets closer to $\lambda_0=1$ as the size of the system grows. 

\begin{figure}[h]
\includegraphics[width=8cm]{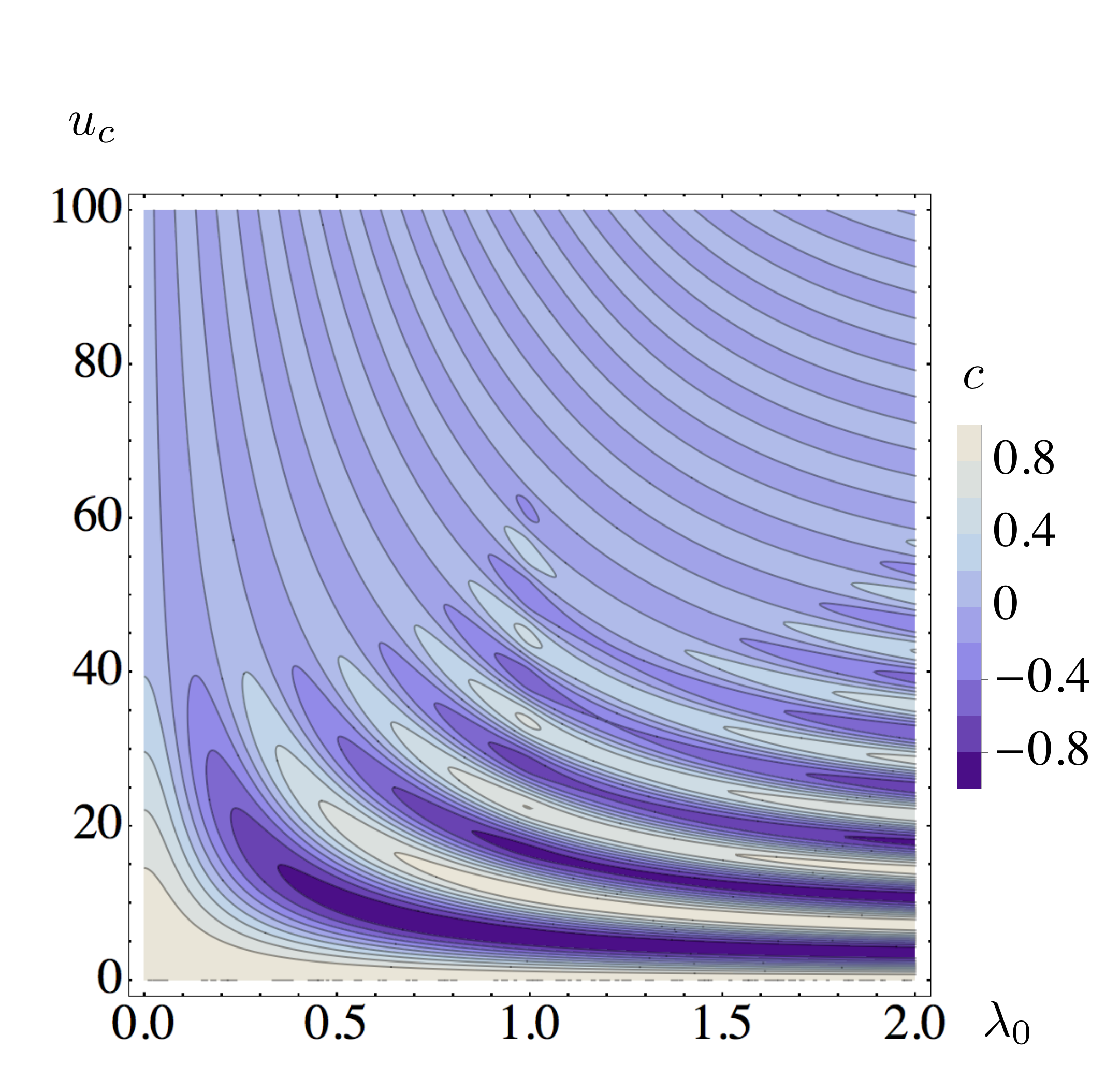}
\caption{(Color online) Contour Plot of the real part of the characteristic function of the work for $N=100$ spins for temperature $\beta=0.1$. The increase of the values of the function goes from the violet (small values) to the white colour (large values).} 
\label{RealChiT} 
\end{figure}

Finally, in Fig.~\ref{RealChiT} we show $\text{Re}[\chi(u)]$ for $N=100$ and $\beta=0.1$ (thus corresponding to a high temperature situation). As already noticed for the plots of the variance and the skewness, also in the characteristic function we see that the signature of the transition is sharp just as long as we stay in the low temperature regime, while the increase of the temperature washes out any evidence of criticality [cf. Fig.~\ref{RealChiN} {\it (a)}] - in line with the expectation that large temperatures would enforce the emergence of classical thermodynamic irreversibility, masking any effects arising from genuine quantum fluctuations~\cite{Dorner}. These features are strong suggestions that also in the characteristic function of the work, done by quenching the transverse field in the Ising model, we can observe a signature of the phase transition.

It is also known that the Loschmidt echo is the modulus square of the work characteristic function~\cite{Silva}. Indeed, the authors of Ref.~\cite{ZanardiEcho} found that, for the case of the transverse Ising model, the Loschmidt echo can be used as a witness of the quantum phase transition. Using a numerical analysis, Ref.~\cite{ZanardiEcho} predicts a non-analytical behaviour of the echo and conjectures 
invariance under the transformation $\Delta \lambda\rightarrow \alpha \Delta \lambda$ and $N\rightarrow N/\alpha $. This is confirmed by our investigation. In fact we observed the very same scaling in our numerical analysis, as shown for example in Fig.~\ref{RealChiN} (a) and Fig.~\ref{RealChiN} (b) obtained respectively for the values $N=100, \Delta \lambda=0.01$ and $N=10, \Delta\lambda=0.1$. Our analysis in the quest for a signature of the phase transition in the work characteristic function can thus be seen as complementary to the one in Ref.~\cite{ZanardiEcho}, albeit based on different analytical tools.

\section{Conclusions}
\label{conc}
We have studied in detail the statistics of the work done on a quantum many-body system by quenching its work parameter. We obtained a simple relation that links the cumulants of the distribution generated by the system magnetization to the susceptibilities of the magnetization itself. This gives a simple physical interpretation to all the cumulants of the work distribution in a special case: a process that involves an observable that commutes with the unperturbed Hamiltonian of the system.
This circumstance highlighted the fact that there are processes for which the whole non-equilibrium thermodynamics can be obtained by simply looking at the statistics of some quantum observables, and so doing single projective measurements, and other processes for which this is not sufficient. Notably one of the consequences of this is clear when we come to study a many-body system with a criticality.
It is in fact non trivial that we should expect to observe signatures of the phase transition in high order moments of the work distribution. In fact we showed that the variance of the work is not solely ascribed to the susceptibility, and the skewness is not proportional to a cumulant of the magnetization distribution. However we observed a signature of the transition in both this quantity and even in the probability distribution itself. Recently a procedure to experimentally measure the characteristic function of the work  by making use of a simple interferometric scheme has been proposed \cite{Oxf}, and used to measure the real part of the characteristic function of the work in a quantum system \cite{Batalhao}. Our results thus suggest the possibility to experimentally observe signatures of quantum phase transitions in systems with criticality by looking at the full statistics of the work.

We have shown that the study of the full statistics of the work in a quantum many-body system, even in the simple case of a sudden quench of the Hamiltonian, is not trivial. In particular, as the work distribution strongly depends upon the structure of the energy levels during the protocol, there could be several physical properties of a quantum many-body system that could be studied by making use of the full statistics of work. The identification of the connection between physical observables and the cumulants of the work in the most general scenario, in addition to helping us in assessing the statistics of work via single projective measurements, could be very important in understanding the emergence of macroscopic thermodynamics from the fully quantum microscopic description of the system. Indeed this topic is one of the main motivations behind the study of thermodynamics and we believe that much work still has to be done in the future in this direction. The role played by the model that we have studied in detail in this paper, the Ising model, as a key benchmark in quantum many body physics makes our study relevant to a widespread realm of disciplines, from condensed matter and solid-state physics to statistical mechanics.

\acknowledgments 
The authors are indebted with T. S. Batalh\~ao, J. Goold, R. Serra, and Peter Talkner for invaluable discussions. This work has been supported by a PERFEST award (LF) from Universit\`{a} degli Studi di Palermo, the Marie Curie Action, the UK EPSRC (EP/G004579/1 and EP/L005026/1), the John Templeton Foundation (grant ID 43467), the EU Collaborative Project TherMiQ (Grant Agreement 618074) and by a Marie Curie Intra European Fellowship within the 7th European Community Framework Programme through the project NeQuFlux grant n. 623085 (MC). TJGA is supported by the European Commission, the European Social Fund and the Region Calabria through the program POR Calabria FSE 2007-2013-Asse IV Capitale Umano-Obiettivo Operativo M2. AX acknowledges funding from the Royal Commission for the Exhibition of 1851. Part of this work was supported by the COST Action MP1209 "Thermodynamics in the quantum regime".


\renewcommand{\theequation}{A-\arabic{equation}}
\setcounter{equation}{0}  
\section*{APPENDIX A}  

\label{a1}
Given the Hamiltonian $\hat{\mathcal H}(\lambda)=\hat{\mathcal H}_{ss}-\lambda \hat{M}_z$, under the hypotheses $[\hat{\mathcal H}_{ss},\hat{M}_z]= 0$, in this appendix we will derive Eq.~\eqref{theoremcumulants}. In this appendix we are concerned with only the thermal state so that we will drop the subscript $G$ used to indicate this state. The average value of the Observable $\hat{M}_z$ in the initial thermal state in given by
\begin{equation}
\langle \hat{M}_z \rangle = \frac{\tr{\hat{M}_z e^{-\beta(\hat{\mathcal H}_{ss}-\lambda_0 \hat{M}_z)}}}{\mathcal{Z}(\lambda_0)}.
\end{equation}
It is easy to show that the average value of $\hat{M}_z^n$, for every finite positive integer $n$, is obtained from the $n^{th}$ order derivative of the moment generating function
\begin{equation}
G (v,\lambda_0)= \frac{\tr{e^{i v \hat{M}_z} e^{-\beta(\hat{\mathcal H}_{ss}-\lambda_0 \hat{M}_z)}}}{\mathcal{Z}(\lambda_0)},
\end{equation}
similarly to the definition of the characteristic function of the work, but with the important difference that here we take just one projective measurement of the observable $\hat{M}_z$. We can define the cumulants of the statistics of $\hat{M}_z$
\begin{equation}
C_{n}=\frac{1}{i^n}\frac{d^n}{d v^n}\log G(v,\lambda_0)\Bigg|_{v=0}.
\end{equation}
Given these definition in this appendix we will show the the validity of the relation
\begin{equation}
\frac{d^n}{d {\lambda_0}^n} \langle \hat{M}_z \rangle=\beta^n C_{n+1}
\label{theorem}
\end{equation}
by mathematical induction on the integer $n$.\\
The validity of Eq. \eqref{theorem} for $n=1$ is the known result that relates the magnetic susceptibility to the variance of the magnetization, which is valid for $[\hat{\mathcal H}_{ss},\hat{M}_z]= 0$. Now we suppose the validity of Eq. \eqref{theorem} for $n$ and see if it is still valid for $n+1$
\begin{equation}
\begin{aligned}
\frac{d^{n+1}}{d {\lambda_0}^{n+1}} \langle \hat{M}_z \rangle&=\beta^n \frac{d}{d {\lambda_0}} C_{n+1}=\beta^n \frac{1}{i^{n+1}}\frac{d^{n+1}}{d v^{n+1}}\left(\frac{d}{d {\lambda_0}}\log G (v,{\lambda_0})\right)\Bigg|_{v=0} 
\end{aligned}
\label{np1}
\end{equation}
So we need to evaluate the last term inside round brackets
\begin{equation}
\begin{aligned}
&\frac{d}{d \lambda_0}\log G (v,\lambda_0)=\frac{d}{d \lambda_0} \log \left( \frac{\tr{e^{i v \hat{M}_z}e^{-\beta(\hat{\mathcal H}_{ss}-\lambda_0 \hat{M}_z)}}}{{\cal Z}(\lambda_0)} \right)\\
&=\frac{1}{G(v,\lambda_0) \mathcal{Z}(\lambda_0)^2}\Biggl(\mathcal{Z}(\lambda_0)\frac{d}{d \lambda_0}\tr{e^{i v \hat{M}_z}e^{-\beta(\hat{\mathcal H}_{ss}-\lambda_0 \hat{M}_z)}}+\\
&-\tr{e^{i v \hat{M}_z}e^{-\beta(\hat{\mathcal H}_{ss}-\lambda_0 \hat{M}_z)}} \frac{d}{d \lambda_0}\mathcal{Z}(\lambda_0)\Biggr)
\end{aligned}
\label{logG}
\end{equation}
In the last equation the two derivatives inside round brackets are given respectively by
\begin{equation}
\frac{d}{d \lambda_0} \tr{e^{i v \hat{M}_z}e^{-\beta(\hat{\mathcal H}_{ss}-\lambda_0 \hat{M}_z)}}=\tr{e^{i v \hat{M}_z}\beta \hat{M}_z e^{-\beta(\hat{\mathcal H}_{ss}-\lambda_0 \hat{M}_z)}}
\label{firstterm}
\end{equation}
and
\begin{equation}
\frac{d}{d\lambda_0}\mathcal{Z}(\lambda_0)=\beta \mathcal{Z}(\lambda_0)\langle \hat{M}_z \rangle.
\label{secondterm}
\end{equation}
Plugging Eq. \eqref{firstterm} and Eq. \eqref{secondterm} into Eq. \eqref{logG} we get
\begin{equation}
\frac{d}{d \lambda_0}\log G(v,\lambda_0)=\beta \left( \frac{1}{i} \frac{d}{dv}\log G(v,\lambda_0) -\langle \hat{M}_z \rangle \right).
\end{equation}
Eventually plugging the previous into Eq. \eqref{np1} we obtain
\begin{equation}
\frac{d^{n+1}}{d \lambda_0^{n+1}} \langle \hat{M}_z \rangle=\beta^{n+1}\frac{1}{i^{n+2}}\frac{d^{n+2}}{d v^{n+2}}\log G(v,\lambda_0)\Bigg|_{v=0}=\beta^{n+1} C_{n+2}.
\end{equation}

\renewcommand{\theequation}{B-\arabic{equation}}
\setcounter{equation}{0}  
\section*{APPENDIX B}  

\label{a2}

In this Appendix we calculate the explicit expression for the $n^{\text{th}}$ moments of the work following a sudden quench of $\lambda$. To achieve this task we need the derivatives of the characteristic function for the sudden quench
\begin{equation}
\partial_u\chi(u,\tau)=i\text{Tr} \bigg[\e{iu\hat{\cal H}(\lambda_\tau)}\bigg(\hat{\cal H}(\lambda_{\tau})-\hat{\cal H}(\lambda_{0})\bigg)\e{-iu\hat{\cal H}(\lambda_0)}\hat{\rho}_0'\bigg].
\label{firstderivative}
\end{equation}
so that for the first moment we get the well-known result
\begin{equation}
\langle W \rangle=\tr{\left( \hat{\cal H}(\lambda_{\tau})-\hat{\cal H}(\lambda_{0}) \right)\hat{\rho}_0'}.
\end{equation}
We now differentiate \eqref{firstderivative} once more to get 

\begin{widetext}
\begin{equation}
\begin{aligned}
\partial^2_u\,\chi(u,\tau)&=-\tr{ \e{iu\hat{\cal H}(\lambda_\tau)}\left\{  \hat{\cal H}(\lambda_{\tau}) \left( \hat{\cal H}(\lambda_{\tau})-\hat{\cal H}(\lambda_{0}) \right) - \left( \hat{\cal H}(\lambda_{\tau})-\hat{\cal H}(\lambda_{0}) \right)\hat{\cal H}(\lambda_0)\right\}\e{-iu\hat{\cal H}(\lambda_0)}\hat{\rho}_0'}\\
&=-\tr{ \e{iu\hat{\cal H}(\lambda_\tau)}\left\{ \hat{\cal H}(\lambda_{\tau})^2-2\hat{\cal H}(\lambda_{\tau})\hat{\cal H}(\lambda_{0})+\hat{\cal H}(\lambda_{0})^2 \right\}\e{-iu\hat{\cal H}(\lambda_0)}\hat{\rho}_0'}.
\end{aligned}
\label{secondsecond}
\end{equation}
\end{widetext}
In general $[\hat{\cal H}(\lambda_{\tau}),\hat{\cal H}(\lambda_{0})] \ne 0$, 
and this makes the term in the curly brackets in Eq.~\eqref{secondsecond} different from $\left( \hat{\cal H}(\lambda_{\tau})-\hat{\cal H}(\lambda_{0}) \right)^2$. Thus when we address the second moment of the work characteristic function we find
\begin{equation}
\langle W^2 \rangle =\tr{\left(\hat{\cal H}(\lambda_\tau)^2-2\hat{\cal H}(\lambda_\tau)\hat{\cal H}(\lambda_0)+\hat{\cal H}(\lambda_0)^2\right)\hat{\rho}_0'}.
\end{equation}
Now using the fact the $[\hat{\rho}_0',\hat{{\cal H}}(\lambda_0)]=0$ ($\hat{\rho}_0'$ is the projected part of $\hat{\rho}_0$ onto the eigenbasis of $\hat{\cal H}(\lambda_0)$) and the cyclic permutation invariance of the trace we get
\begin{equation}
\langle W^2 \rangle=\tr{\left( \hat{\cal H}(\lambda_{\tau})-\hat{\cal H}(\lambda_{0}) \right)^2\hat{\rho}_0'}.
\end{equation}
However, for the higher moments we have
\begin{equation}
\langle W^n \rangle\ne\tr{\left( \hat{\cal H}(\lambda_{\tau})-\hat{\cal H}(\lambda_{0}) \right)^n\hat{\rho}_0'}~~\forall n>2.
\end{equation}
By extending the approach used in order to obtain Eq.~\eqref{secondsecond}, further, it is straightforward to see that the $n$-th moment of the work characteristic function can be written as
\begin{equation}
\langle W^n \rangle = 
\tr{\sum_{k=0}^n \binom{n}{k} \hat{\cal H}(\lambda_{\tau})^{(n-k)}\hat{\cal H}(\lambda_{0})^{k}(-1)^k \hat{\rho}_0'}
\end{equation}
for any finite value of $n$ and for any initial state $\hat{\rho}_0'$.

\renewcommand{\theequation}{C-\arabic{equation}}
\setcounter{equation}{0}  
\section*{APPENDIX C}  

\label{a3}

In this appendix we will show that, if we are in the case $[\hat{M}_z,{\cal H}_{ss}] \ne 0$, the relation
\begin{equation}
\chi_M = \beta \Delta \hat{M}_z^2+\widetilde{\chi}_M
\label{susceNEW}
\end{equation}
holds, and we will find an explicit expression for $\widetilde{\chi}_M$. The Hamiltonian under scrutiny reads $\hat{\cal H}(\lambda)=\hat{\cal H}_{ss}-\lambda \hat{M}_z$. According to the definition of susceptibility, we have 
\begin{equation}
\label{chim}
\begin{aligned}
\chi_M &:= \frac{d \langle \hat{M}_z \rangle}{d \lambda_0} = \frac{1}{{\cal Z}(\lambda_0)^2 }\Biggl( {\cal Z}(\lambda_0)\,\partial_{\lambda_0}\tr{\hat{M}_z e^{-\beta(\hat{\cal H}_{ss} -\lambda_0 \hat{M}_z)}}\\
&-\tr{\hat{M}_z e^{-\beta(\hat{\cal H}_{ss} -\lambda_0 \hat{M}_z)}} \partial_{\lambda_0}{\cal Z}(\lambda_0)\Biggr).
\end{aligned}
\end{equation}
We thus first need to calculate the derivative of the partition function with respect to $\lambda_0$, which can be cast into the form
\begin{equation}
\begin{aligned}
\partial_{\lambda_0} {\cal Z}
=\partial_{\lambda_0} \tr{\hat{\openone}-\beta(\hat{\cal H}_{ss}-\lambda_0 \hat{M}_z)+\frac{(-\beta)^2}{2!}(\hat{\cal H}_{ss}-\lambda_0 \hat{M}_z)^2+...}
\end{aligned}
\end{equation}
Although we can invert the order of tracing and differentiating, we must pay attention to the non commutativity of the operators. By using the cyclic permutation property inside the trace, a straightforward calculation leads us to 
\begin{equation}
\begin{aligned}
&\tr{\partial_{\lambda_0}(\hat{\cal H}_{ss}-\lambda_0 \hat{M}_z)^n}
=\tr{-n \hat{M}_z (\hat{\cal H}_{ss}-\lambda_0 \hat{M}_z)^{n-1}},
\end{aligned}
\end{equation}
which in turn gives us 
$\tr{\partial_{\lambda_0} e^{-\beta(\hat{\cal H}_{ss}-\lambda_0 \hat{M}_z)}}=\tr{\beta \hat{M}_z e^{-\beta (\hat{\cal H}_{ss}-\lambda_0 \hat{M}_z)}}$ and finally
\begin{equation}
\partial_{\lambda_0} {\cal Z}=\beta {\cal Z}(\lambda_0) \langle \hat{M}_z \rangle,
\end{equation}
which is exactly the relation in Eq.~\eqref{secondterm} that we have now proven to be valid also for $[\hat{M}_z,{\cal H}_{ss}] \ne 0$.
A very similar calculation leads us to
\begin{equation}
\begin{aligned}
&\tr{\hat{M}_z \partial_{\lambda_0}(\hat{H}_{ss}-\lambda_0 \hat{M}_z)^{n} }=\\
&-\tr{\sum_{k=0}^{n-1}\hat{M}_z(\hat{H}_{ss}-\lambda_0 \hat{M}_z)^k \hat{M}_z (\hat{H}_{ss}-\lambda_0 \hat{M}_z)^{n-k-1}},
\end{aligned}
\end{equation}
which can be used to obtain 
\begin{equation}
\begin{aligned}
&\partial_{\lambda_0} \tr{\hat{M}_z e^{-\beta(\hat{H}_{ss}-\lambda_0 \hat{M}_z)} }=\tr{\beta \hat{M}_z^2 e^{-\beta(\hat{H}_{ss}-\lambda_0 \hat{M}_z)} }\\
+&\tr{\sum_{n=1}^\infty \sum_{k=0}^{n-1} \frac{(-\beta)^n}{n!} \left[(\hat{H}_{ss}-\lambda_0 \hat{M}_z)^k,\hat{M}_z\right] \hat{M}_z \left(\hat{H}_{ss}-\lambda_0 \hat{M}_z\right)^{n-k-1}}.
\end{aligned}
\end{equation}
Therefore, for $[\hat{M}_z,{\cal H}_{ss}] \ne 0$ the relation in Eq.~\eqref{susceNEW} holds with the correction term given by
\begin{equation}
\begin{aligned}
&\widetilde{\chi}_M=\frac{1}{{\cal Z}(\lambda_0)}\times\\
&\tr{\sum_{n=1}^\infty \sum_{k=0}^{n-1} \frac{(-\beta)^n}{n!} \left[(\hat{H}_{ss}-\lambda_0 \hat{M}_z)^k,\hat{M}_z\right] \hat{M}_z \left(\hat{H}_{ss}-\lambda_0 \hat{M}_z\right)^{n-k-1}}
\end{aligned}
\end{equation}

\renewcommand{\theequation}{D-\arabic{equation}}
\setcounter{equation}{0}  
\section*{APPENDIX D}  

\label{a4}

In this appendix we diagonalise the model
\begin{equation}
\hat{{\cal H}}(\lambda)=-\sum_{i=1}^N \hat{\sigma}_i^x\hat{\sigma}_{i+1}^x -\lambda \sum_{i=1}^N \hat{\sigma}_i^z,
\end{equation}
by mapping the spin operators into spinless fermionic ones defined as 
\begin{equation}
\label{JW}
\hat{c}_j=\frac{1}{2}\prod_{l=1}^{j-1}\hat{\sigma}_l^z(\hat{\sigma}_j^x+i\hat{\sigma}_j^y),\,~~\hat{c}_j^{\dagger}=\frac{1}{2}\prod_{l=1}^{j-1}\hat{\sigma}_l^z(\hat{\sigma}_j^x-i\hat{\sigma}_j^y).
\end{equation}
We can define the parity operators
\begin{equation}
\hat{P}^{\pm}=\frac{1}{2}\biggl[1\pm \prod_{j=1}^N (1-2\hat{c}_j^{\dagger}\hat{c}_j)\biggr]
\end{equation}
which are projectors on subspaces with even ($P^+$) and odd ($P^-$) numbers of c-quasiparticles and $H^{\pm}$ are the corresponding reduced Hamiltonians
\begin{equation}
\hat{H}=\hat{P}^+ \hat{H}^+ \hat{P}^+ + \hat{P}^- \hat{H}^- \hat{P}^-.
\end{equation}
The only difference between $H^+$ and $H^-$ is that in $H^+$ we impose the anti-periodic boundary condition $\hat{c}_{N+1} = -\hat{c}_1$ and in $H^-$ we impose the periodic boundary condition $\hat{c}_{N+1} = \hat{c}_1$. The parity of the chain is a good quantum number so that the dynamics does not mix the two parity subspaces. The state we deal with is a thermal state, so that in principle we would need to take both subspaces into account. However we are interested in the thermodynamical limit and in this limit it is known that the results are exact also considering only one parity projection of the Hamiltonian. That is why in general in the paper we make the identification $\hat{H}=\hat{H}^+$ and we do not distinguish between them anymore.
The following step in the diagonalisation is a Fourier transformation, which is accomplished by
\begin{equation}
\hat{c}_j=\frac{e^{-i\pi/4}}{\sqrt{N}} \sum_{k \in K^+} \hat{c}_k e^{i k_j}
\end{equation}
with $K^+=\{k=\pm\pi(2n-1)/n\}$ and $n=1,...,{N}/{2}$, as we are restricting our attention to the even parity subspace of the model. Then we apply the Bogolioubov transformation
\begin{equation}
\hat{c}_{\pm k}=\hat{\gamma}_{\pm k} \cos(\phi_k/2)\mp\hat{\gamma}_{\mp k}^{\dagger}\sin(\phi_k/2)
\label{bogo}
\end{equation}
with the Bogolioubov angles defined as
\begin{equation}
\begin{aligned}
\cos(\phi_k)&=\frac{\lambda-\cos(k)}{\sqrt{\sin^2 (k) + [\lambda-\cos(k)]^2}}\\
\sin(\phi_k)&=\frac{\sin(k)}{\sqrt{\sin^2 (k) + [\lambda-\cos(k)]^2}},
\end{aligned}
\end{equation}
and $\{\hat{\gamma}_k,\hat{\gamma}_k^{\dagger} \}$ is a set of fermionic operators.
With this notation, the diagonal form of the \emph{pre-quench} Hamiltonian reads
\begin{equation}
\hat{\cal H}(\lambda_0)=\sum_{k \in K^+}\epsilon_k(\lambda_0) \biggl( \hat{\gamma}_k^{\dagger}\hat{\gamma}_k-\frac{1}{2} \biggr)
\label{diagonalHpre}
\end{equation}
with the dispersion relation
\begin{equation}
\epsilon_k(\lambda)=2\sqrt{\sin^2 (k) +[\lambda-\cos(k)]^2}.
\end{equation}
Following an analogous approach, the \emph{post-quench} Hamiltonian is found to be given by the diagonal model
\begin{equation}
\hat{\cal H}(\lambda_{\tau})=\sum_{k \in K^+}\epsilon_k(\lambda_{\tau}) \biggl( \hat{\gamma}_k'^{\dagger}\hat{\gamma}'_k-\frac{1}{2} \biggr)
\label{diagonalHpost}
\end{equation}
obtained from Eq.~\eqref{diagonalHpre} with $\lambda_0\rightarrow \lambda_{\tau}$ and $\hat{\gamma}_k\rightarrow\hat{\gamma}'_k$.
The characteristic function for this system is obtained by evaluating the trace in Eq.~\eqref{characteristicfunctionsudden} over the eigenstates of the initial Hamiltonian. So we need to express the post-quench Hamiltonian eigenbasis in terms of the pre-quench eigenbasis. To this aim we need first to connect the two classes of fermionic operators. This is done by simply inverting Eq.~\eqref{bogo} for both the pre- and post-quench fermionic operators obtaining the relations 
\begin{equation}
\begin{aligned}
\hat{\gamma}'_k &=\hat{\gamma}_k \cos(\Delta_k/2)+\hat{\gamma}_{-k}^{\dagger}\sin (\Delta_k/2)\\
\hat{\gamma}'_{-k} &=\hat{\gamma}_{-k} \cos(\Delta_k/2)-\hat{\gamma}_{k}^{\dagger}\sin (\Delta_k/2).
\end{aligned}
\end{equation}
and $\Delta_k=\tilde{\phi}_k-\phi_k$ is the difference in the pre- and post-quench Bogoliubov angles. The relation between the two vacuum states is
\begin{equation}
| 0_k, 0_{-k}\rangle=\left( \cos \left(\frac{\Delta_k}{2}\right) + \sin \left(\frac{\Delta_k}{2}\right) \hat{\gamma}_k'^{\dagger} \hat{\gamma}_{-k}'^{\dagger}\right) |0_k',0_{-k}'\rangle.
\end{equation}

%

\end{document}